\documentclass[a4paper]{article}
\usepackage{a4wide}
\usepackage[UKenglish]{babel}
\usepackage{amsmath,amssymb,amsthm,mathtools,empheq,mathrsfs}

\usepackage{bbm}
\usepackage{graphicx,subfigure}
    \graphicspath{{./}{fig/}}
\usepackage{xcolor}
\usepackage[colorlinks=true,linkcolor=blue,citecolor=red]{hyperref}
\usepackage{comment}
\usepackage{authblk}
\usepackage{accents}
\usepackage[inline]{enumitem}

\usepackage{tikz}
    \usetikzlibrary{positioning}
    \usetikzlibrary{arrows}
    \usetikzlibrary{arrows.meta}
    \usetikzlibrary{calc}

\newcommand{\beq}{\begin{equation}}
\newcommand{\ddt}{\frac{{\rm d}}{{\rm d}t}}
\newcommand{\di}{{\rm d}}
\newcommand{\eeq}{\end{equation}}

\newcommand{\bv}{\mathbf{v}}
\newcommand*\tcircle[1]{%
  \raisebox{-0.5pt}{%
    \textcircled{\fontsize{7pt}{0}\fontfamily{phv}\selectfont #1}%
  }%
}
\newcommand{\ubar}[1]{\underaccent{\bar}{#1}}

\def\red{\textcolor{black}}

\allowdisplaybreaks

\title{
Heterogeneously structured compartmental models of epidemiological systems: from individual-level processes to population-scale dynamics\thanks{T.L., M.S., and A.T. gratefully acknowledge support from the Italian Ministry of University and Research (MUR) through the PRIN 2020 project (No. 2020JLWP23) ``Integrated Mathematical Approaches to Socio–Epidemiological Dynamics'' (CUP: E15F21005420006).}}

\author{Emanuele Bernardi}
\author{Tommaso Lorenzi}
\author{Mattia Sensi}
\author{Andrea Tosin}
\affil{{\small Department of Mathematical Sciences ``G. L. Lagrange'' \\ Politecnico di Torino, Italy}}
\date{}

\begin{document}
\maketitle

\begin{abstract}
We develop a general modelling framework for compartmental epidemiological systems structured by continuous variables which are linked to the levels of expression of compartment-specific traits. We start by formulating an individual-based model that describes the dynamics of single individuals in terms of stochastic processes. Then we formally derive: (i) the mesoscopic counterpart of this model, which is formulated as a system of integro-differential equations for the distributions of individuals over the structuring-variable domains of the different compartments; (ii) the corresponding macroscopic model, which takes the form of a system of ordinary differential equations for the fractions of individuals in the different compartments and the mean levels of expression of the traits represented by the structuring variables. We employ a reduced version of the macroscopic model to obtain a general formula for the basic reproduction number, $\mathcal{R}_0$, in terms of key parameters and functions of the underlying microscopic model, so as to illustrate how such a modelling framework makes it possible to draw connections between fundamental individual-level processes and population-scale dynamics. Finally we apply the modelling framework to case studies based on classical compartmental epidemiological systems, for each of which we report on Monte Carlo simulations of the individual-based model as well as on analytical results and numerical solutions of the macroscopic model. 
\end{abstract}

\section{Introduction}
\begin{figure}[ht]
\centering
\subfigure[]{
    \begin{tikzpicture}
	\node at (0,0) {\includegraphics[width=.3\linewidth]{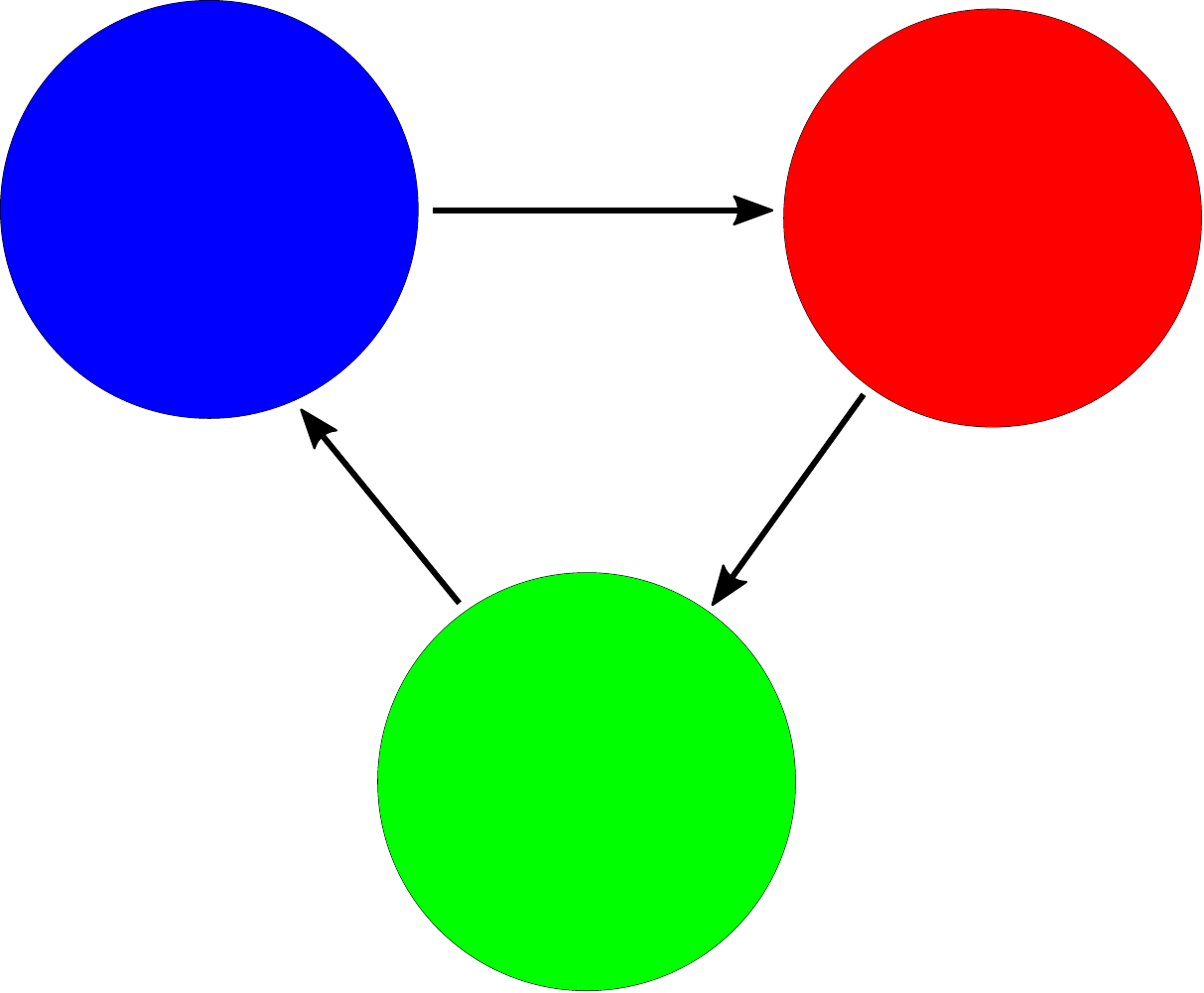}};
        \node at (-2,2.25) {\Large $S$};
        \node at (2,2.25) {\Large $I$};
        \node at (0,-2.35) {\Large $R$};
    \end{tikzpicture}
} \qquad\qquad
\subfigure[]{
    \begin{tikzpicture}
        \node at (0,0) 	{\includegraphics[width=.3\linewidth]{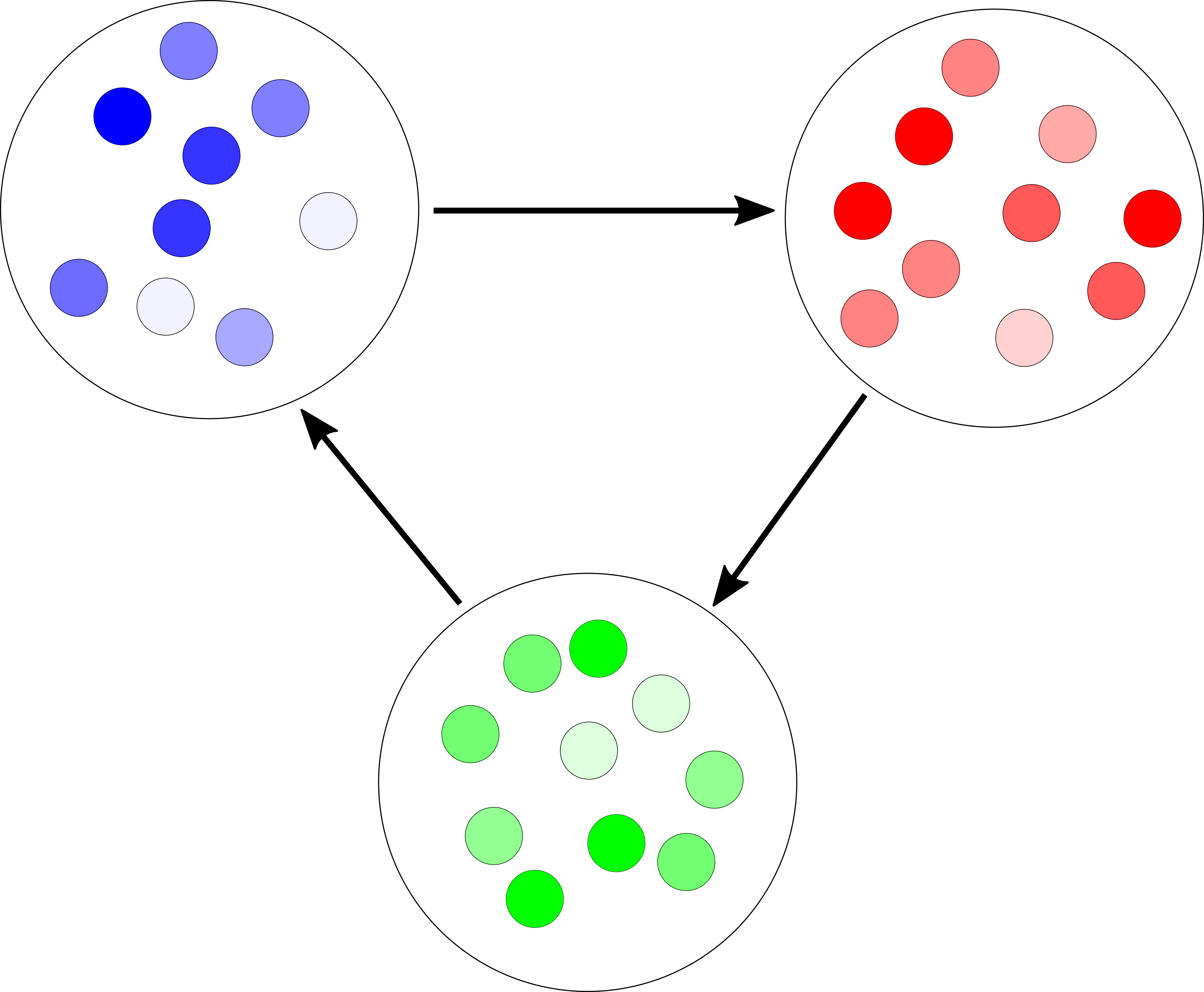}};
        \node at (-2,2.25) {\Large $S$};
        \node at (2,2.25) {\Large $I$};
        \node at (0,-2.35) {\Large $R$};
    \end{tikzpicture}
}
\caption{The schematic in panel (a) shows the flow diagram for a homogeneous SIRS system, where individuals in each compartment are assumed to be identical, while the schematic in panel (b) shows the corresponding flow diagram for a heterogeneous SIRS system, where individuals in each compartment express a specific characteristic to different degrees, which are here represented by different colour shades
}
\label{fig:intro}
\end{figure}
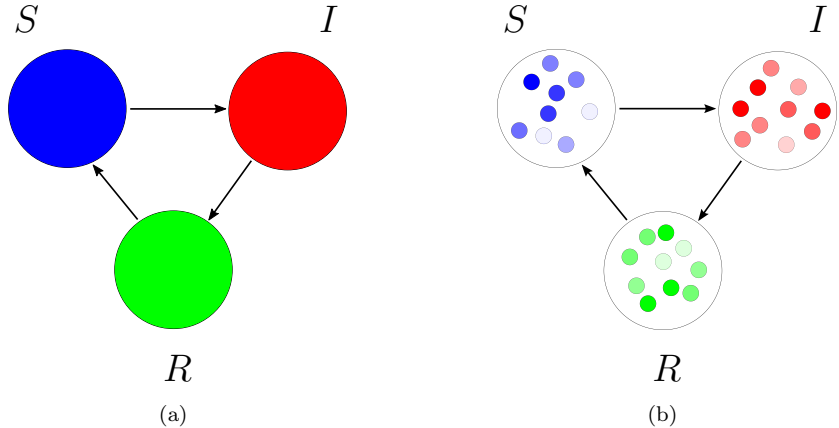

The basic tenet of compartmental models of epidemiological systems formulated as ordinary differential equations (ODEs) is that each compartment is homogeneous (i.e. it is composed of individuals that are identical)~\cite{brauer2008compartmental,hethcote1989three,hethcote2000mathematics}. For instance, focusing on SIRS systems~\cite{bulai2024geometric,jardon2021geometric,jardon2021geometric2,kaklamanos2024geometric}, all susceptible individuals are assumed to have the same level of resistance to infection, all infectious individuals express the same viral load, and all recovered individuals share the same immunity level (cf. the schematic in Figure~\ref{fig:intro}(a)). However, it has long been recognised that individuals in the same compartment are rarely (if ever) homogeneous -- i.e. still focusing on SIRS systems, there is variability in the level of resistance to infection, viral load, and immunity level amongst individuals (cf. the schematic in Figure~\ref{fig:intro}(b)).

An increasing body of literature provides compelling evidence for the fact that this form of inter-individual heterogeneity plays a pivotal role in shaping the evolution of epidemic processes~\cite{britton2020mathematical,chabas2018evolutionary,elie2022source,jones2021viral,tkachenko2021time,vanderwaal2016heterogeneity,woolhouse1997heterogeneities,yates2006pathogen}. As a result, continuously structured compartmental models formulated as partial differential equations and integro-differential equations (IDEs) have also been developed, wherein one or more compartments are structured by continuous variables that capture variation in relevant characteristics amongst individuals, such as age~\cite{busenberg1991global,inaba1990threshold,inaba2017age,thieme2009spectral}, resistance to infection or immunity level~\cite{barbarossa2015immuno,gandolfi2015epidemic,iacono2012evolution,lorenzi2023modelling,lorenzi2021evolutionary}, viral or pathogen load ~\cite{banerjee2020immuno,della2022sir,della2023sir,gandolfi2015epidemic,lorenzi2023modelling,loy2021viral}, and other phenotypic~\cite{abi2021asymptotic,almeida2021final,burie2020asymptotic,burie2020slow,burie2020concentration,djidjou2017steady,karev2019trait,novozhilov2012epidemiological,veliov2005effect} or socio-economic~\cite{bernardi2022effects,dimarco2020wealth,dimarco2021kinetic,zanella2022kinetic} characteristics. 

Building upon previous work on continuously structured compartmental models of epidemiological systems~\cite{della2022sir,della2023sir}, in this paper we generalise the modelling approach for SI systems that we presented in~\cite{lorenzi2023modelling} by developing a modelling framework that comprises an arbitrary number of compartments, each structured by a continuous variable that is linked to the level of expression of some compartment-specific traits. 

\begin{figure}[ht]
	\centering
	\begin{tikzpicture}
		\node at (0,0) 	{\includegraphics[width=0.6\linewidth]{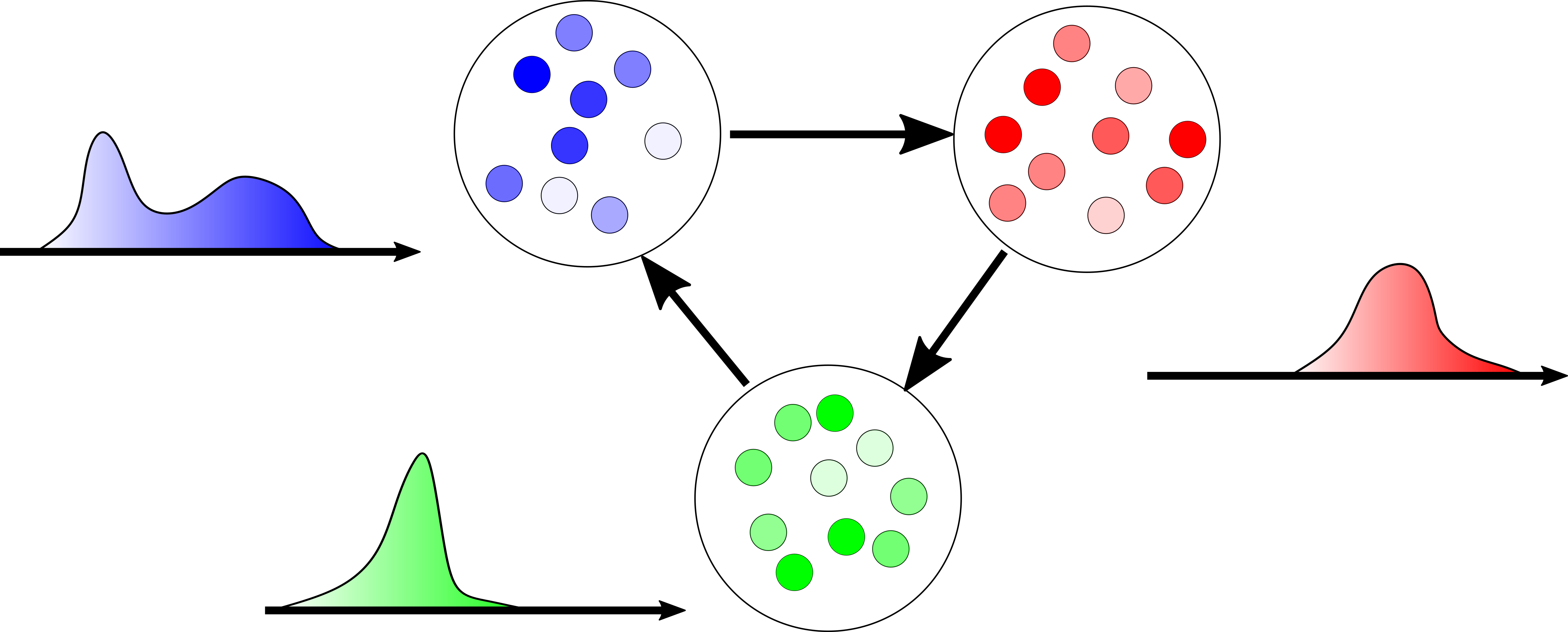}};
		\node at (-1.7,2.15) {\Large $S$};
		\node at (2.2,2.15) {\Large $I$};
		\node at (0.2,-2.25) {\Large $R$};
		\node at (-3.5,1.35) {$n_S(t,v_S)$};
		\node at (-3.5,0) {$\mathcal{V}_S$};
		\node at (4.5,0.2) {$n_I(t,v_I)$};
		\node at (3.5,-0.7) {$\mathcal{V}_I$};
		\node at (-3.4,-1.25) {$n_R(t,v_R)$};
		\node at (-2.2,-2.1) {$\mathcal{V}_R$};
	\end{tikzpicture}
		\caption{Schematic showing the flow diagram for a heterogeneous SIRS system along with the related representation in terms of the population density functions, $n_i(t,v_i)$, each representing the distribution of individuals in a given compartment, labelled by the index $i \in \mathcal{I}:=\{ S, I, R\}$, over the domain of the corresponding compartment-specific structuring-variable, $v_i \in \mathcal{V}_i$, at time $t$
		\label{fig:meso}}
\end{figure}
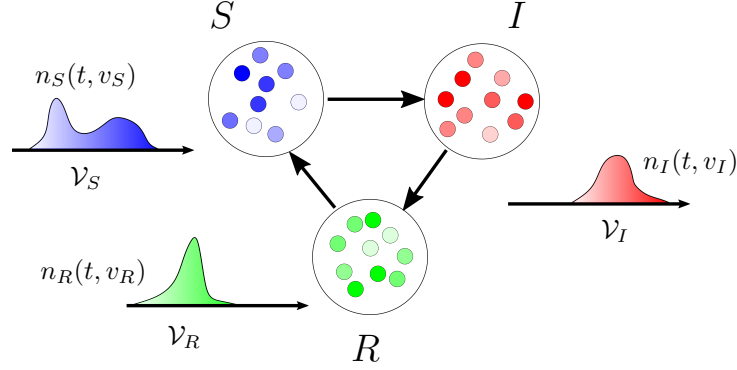

We start by formulating a general continuously structured individual-based model for compartmental epidemiological systems, which describes the dynamics of single individuals in terms of stochastic processes. This model takes into account the effects of both changes in the individuals' level of expression of compartment-specific traits and transitions of individuals between different compartments. Next, we formally derive the mesoscopic counterpart of this model, which is formulated as a system of IDEs for the population density functions that represent the distributions of individuals over the structuring-variable domains of the different compartments (cf. the schematic in Figure~\ref{fig:meso}). Then, considering an appropriately rescaled version of this IDE system, we carry out a formal derivation of the corresponding macroscopic model, which takes the form of a system of ODEs for the fractions of individuals in the different compartments and the mean levels of expression of the traits represented by the structuring variables. 

\red{In summary, the modelling approach presented here provides a coherent multiscale representation, from individual-level processes to population-scale behaviours, of co-evolutionary dynamics in compartmental epidemiological systems that are structured by compartment-specific continuous variables. This is in itself a novelty aspect of our work. In fact, while multiscale modelling of compartmental epidemiological systems has been an active area of research for years now, attention has mainly been focused on unstructured systems, systems where structuring variables are discrete, and systems where structuring variables are continuous but they are the same for all compartments. Instead, here we consider the case where different compartments are structured by different continuous variables, proposing a modelling approach that is flexible enough to be applicable to a wide range of epidemiological systems.}

The remainder of the paper is organised as follows. In Section~\ref{sec:micro} we present the individual-based model, the mesoscopic and macroscopic counterparts of which are then formally derived in Sections~\ref{sec:meso} and~\ref{sec:macro}, respectively. In Section~\ref{sec:examples}, we apply the modelling framework to case studies based on classical compartmental epidemiological systems. For each case study, we report on the results of Monte Carlo simulations of the individual-based model and integrate them with analytical results and numerical solutions of the macroscopic model. In Section~\ref{sec:concl} we conclude with a discussion and propose some future research directions.

\section{Stochastic individual-based model}
\label{sec:micro}
We consider a system of $N>1$ interacting compartments labelled by an index $i\in\mathcal{I}:=\{1,\,\dots,\,N\}$. Each compartment is structured by a specific continuous variable $v_i \in \mathcal{V}_i \subset \mathbb{R}$, where $\mathcal{V}_i$ is a bounded interval. At time $t \in [0,\,\infty)$, we describe the microscopic state of individuals in the system, regarded as being indistinguishable, by the random vector $(I_{t},\,{\bf V}_{t})$, where $I_{t}\in\mathcal{I}$ and ${\bf V}_{t} \equiv (V_{1,t},\,\ldots,\,V_{N,t})\in\mathcal{V}:=\mathcal{V}_1 \times \ldots \times \mathcal{V}_N \subseteq \mathbb{R}^N$. The discrete random variable $I_{t}$ specifies the compartment of an individual at time $t$ (i.e. individuals belonging to compartment $i$ at time $t$ will have $I_{t} = i$), while the continuous random variable $V_{i,t}$ models the level of expression of the trait represented by the $i^{th}$ structuring variable at time $t$.

We let
\beq
\label{eq:f}
(I_{t},\,{\bf V}_{t}) \sim f(t,i,\bv),
\eeq
where \red{we use the symbol $\sim$ to indicate that a random variable is distributed according to a certain probability distribution.} The distribution function $f :  [0,\,\infty)  \times \mathcal{I} \times \mathcal{V} \to \mathbb{R}_+$, being a probability density function, is such that 
\beq
\label{ass:f}
\sum_{i \in \mathcal{I}} \int_{\mathcal{V}} f(t,i,\bv) \di \bv = 1 \;\; \forall \, t \in [0,\infty).
\eeq
We focus on a scenario where between time $t$ and time $t + \Delta t$, with $\Delta t > 0$:
\begin{itemize}
\item[(${\rm I}$)] At rate $\theta_i \in \mathbb{R}^\ast_+$, {\it structuring-variable switching} may occur, whereby an individual in compartment $i$ with expression level $v_i$ of the trait represented by the structuring variable may spontaneously acquire a new expression level $v'_i$, with probability $K_i(v_i'|v_i)$.
\item[(${\rm II}$)] At rate $\zeta \in \mathbb{R}^\ast_+$, {\it compartment switching} may take place, whereby an individual in compartment $i$ with expression level $v_i$ of the trait represented by the structuring variable may transition into compartment $j$. Compartment switching can either be spontaneous (i.e. not driven by interactions with other individuals) with probability $r \in [0,\,1]$, or driven by intra-/inter-compartment interactions (i.e. interactions between individuals of the same compartment/different compartments) with probability $1-r$. In more detail: 
\begin{itemize}
\item[(${\rm IIa}$)] when {\it spontaneous compartment switching} occurs, the individual transitions into compartment $j$, with probability $p(j|i,v_i)$, and acquires a value $v''_{j}$ of the structuring variable of the new compartment, with probability $P(v_j''|j,i,v_i)$\red{. This is the case, for example, of the switch $I \to R$ in the SIRWS model presented in Section \ref{sec:SIRWS_ex}, whereby an infected individual spontaneously recovers and becomes immune};
\item[(${\rm IIb}$)] when {\it interaction-driven compartment switching} occurs, the interaction with an individual in compartment $k$ with value $v_k^\ast$ of the structuring variable leads the individual to transition into compartment $j$, with probability $q(j|i,v_i,k,v_k^\ast)$, and acquire value $v''_{j}$ of the structuring variable of the new compartment, with probability $Q(v_j''|j,i,v_i,k,v_k^\ast)$\red{. Considering again the SIRWS model presented in Section \ref{sec:SIRWS_ex}, this is the case, for example, of: the switch $S+I \to I+I$, whereby a susceptible individual interacts with an infected individual and contracts the disease (i.e. $k=j$); or the switch $W+I \to R+I$, whereby an individual with waning immunity interacts with an infected individual and receives a boost of immunity, becoming recovered and thus immune (i.e. $k\neq j$)}.
\end{itemize}
\end{itemize}
The aforementioned kernels $K_i$, $P$, and $Q$ and the functions $p$ and $q$ satisfy the following assumptions for all $i,j,k \in \mathcal{I}$: 
\begin{align*}
    & K_i(\cdot|\cdot) \in \mathscr{P}(\mathcal{V}_i;C(\mathcal{V}_i)),\quad P(\cdot|j,i,\cdot)\in \mathscr{P}(\mathcal{V}_j;C(\mathcal{V}_i)), \quad p(j|i,\cdot)\in C(\mathcal{V}_i), \\[5pt]
    & Q(\cdot|j,i,\cdot,k,\cdot) \in \mathscr{P}(\mathcal{V}_j;C(\mathcal{V}_i \times \mathcal{V}_k)),\quad q(j|i,\cdot,k,\cdot)\in C(\mathcal{V}_i \times \mathcal{V}_k),
\end{align*}

\begin{subequations}
\begin{align}
    & \text{Supp}(K_i) \subseteq \mathcal{V}_i\times \mathcal{V}_i,\quad K_i \geq 0 \text{ a.e.}, \label{eq:Ki_supp} \\[5pt]
    & \int_{\mathcal{V}_i} K_i(v_i'|v_i) \di v_i' =1 \; \text{and} \; \int_{\mathcal{V}_i} v_i' K_i(v_i'|v_i) \di v_i' \in \mathcal{V}_i \quad
        \forall v_i \in \mathcal{V}_i, \label{eq:Ki_int1} \\[5pt]
    & \int_{\mathcal{V}_i}(v_i')^2 K_{i}(v_i' | v_i) \di v_i' - \left(\int_{\mathcal{V}_i} v_i' K_{i}(v_i' | v_i) \di v_i' \right)^2 > 0 \;\; \forall \, v_i \in  \mathcal{ \mathring V}_i,
    \label{eq:Ki_var}
\end{align}
\end{subequations}

\begin{align*}
    &\text{Supp}(P(\cdot|j,i,\cdot))\subseteq\mathcal{V}_j\times\mathcal{V}_i, \quad
        P(\cdot|j,i,\cdot)\geq 0 \text{ a.e.}, \\[5pt]
    &\int_{\mathcal{V}_j}P(v_j''|j,i,v_i)\di v_j''=1,\;\;\int_{\mathcal{V}_j}v_j''P(v_j''|j,i,v_i)\di v_j''\in\mathcal{V}_j
        \quad\forall v_i\in\mathcal{V}_i,
\end{align*}

\begin{align*}
	& \text{Supp}(Q(\cdot|j,i,\cdot,k,\cdot)) \subseteq \mathcal{V}_j\times \mathcal{V}_i\times \mathcal{V}_k,\quad Q(\cdot|j,i,\cdot,k,\cdot) \geq 0 \text{ a.e.}, \\[5pt]
	& \int_{\mathcal{V}_j} Q(v_j''|j,i,v_i,k,v_k^\ast) \di v_j'' =1, \;\; \int_{\mathcal{V}_j} v_j''Q(v_j''|j,i,v_i,k,v_k^\ast) \di v_j'' \in \mathcal{V}_j \quad \forall (v_i,v_k) \in \mathcal{V}_i\times \mathcal{V}_k,
\end{align*}

\begin{subequations}\label{eq:ker_assump5}
\begin{empheq}[]{align}
	& 	\text{Supp}(p(j|i,\cdot)) \subseteq \mathcal{V}_i, \quad 0\leq p(j|i,\cdot) \leq 1, \nonumber \\[5pt]
	& \text{Supp}(q(j|i,\cdot,k,\cdot)) \subseteq \mathcal{V}_i\times \mathcal{V}_k, \quad 0\leq q(j|i,\cdot,k,\cdot) \leq 1, \nonumber \\[5pt]
        &  		\sum_{j \in \mathcal{I}} p(j|i,v_i) = 1 \;\; \forall v_i \in \mathcal{V}_i,\label{eq:ker_assump5c}
             \\[5pt]
        &  \sum_{\substack{j \in \mathcal{I}}} q(j|i,v_i,k,v_k^\ast)=1 \;\; \forall (v_i,v_k^\ast)  \in \mathcal{V}_i \times \mathcal{V}_k,\label{eq:ker_assump5d}
\end{empheq}
\end{subequations}
where \red{$\mathscr{P}(\mathcal{V}_i)$ denotes the set of probability distributions defined on the measurable space $\left(\mathcal{V}_i,\mathcal{B}\right)$, with $\mathcal{B}$ the Borel $\sigma$-algebra, $C(\mathcal{V}_i)$ denotes the set of continuous functions defined on $\mathcal{V}_i$,} and $\text{Supp}(\cdot)$ denotes the support of $(\cdot)$. Note that the normalisation conditions~\eqref{eq:ker_assump5c} and~\eqref{eq:ker_assump5d} imply that
\begin{subequations}
\begin{empheq}[]{align}
	& 	p(i|i,v_i) = 1- \sum_{\substack{j \in \mathcal{I}\\ j\neq i}} p(j|i,v_i)\;\; \forall v_i \in \mathcal{V}_i,\label{eq:ker_assump6a}
  \\[5pt]
	& q(i|i,v_i,k,v_k^\ast) =  1-\sum_{\substack{j \in \mathcal{I}\\ j\neq i}} q(j|i,v_i,k,v_k^\ast) \;\; \forall (v_i,v_k^\ast)  \in \mathcal{V}_i \times \mathcal{V}_k.\label{eq:ker_assump6b}
\end{empheq}
\end{subequations}

Under this scenario, the evolution of the microscopic state $(I_{t},\,{\bf V}_{t})$ is governed by the following system
\beq
\label{eq:microdynlaws}
\begin{cases}
I_{t + \Delta t} = (1 - Z) \, I_t + Z \,\left[ \Xi I'_t +(1-\Xi) I_t'' \right],
\\\\
V_{i, t + \Delta t} = (1 - Z) \Big[(1 - \Theta_i) \, V_{i, t} + \Theta_i \, V'_{i, t} \Big] + Z \, V''_{i, t}, \quad i\in \mathcal{I},
\end{cases}
\eeq
where $Z$, $\Xi$, and $\Theta_1$, $\ldots$, $\Theta_N$ are independent Bernoulli random variables with parameters $\zeta \Delta t$, $r$, and $\theta_1 \Delta t, \ldots, \theta_N \Delta t$, respectively. Note that we are implicitly assuming $\Delta t$ to be small enough so that $\max \left(\zeta, \theta_1, \ldots, \theta_N\right) \Delta t \le 1$. In the system~\eqref{eq:microdynlaws}, the random variables $I'_t \in \mathcal{I}$ and $I''_t \in \mathcal{I}$ model, respectively, the index of the new compartment an individual belongs to if spontaneous and interaction-driven compartment switching occurs, while the random variable $V''_{i, t} \in \mathcal{V}_i$ models the corresponding value of the structuring variable that is acquired by the individual. Moreover, the random variable $V'_{i, t} \in \mathcal{V}_i$ models the new value of the structuring variable that is acquired by an individual in compartment $i$ if structuring-variable switching occurs.

For ease of presentation, in the remainder of the paper we use the notation 
\beq
\label{eq:hats}
^i\hat{\mathcal{V}} := \mathcal{V}_1 \times \ldots \times \mathcal{V}_{i-1} \times \mathcal{V}_{i+1} \times \ldots \times \mathcal{V}_N, \quad  ^i\hat{\bv} \equiv (v_1, \ldots, v_{i-1}, v_{i+1}, \ldots, v_N).
\eeq 

To incorporate (${\rm I}$) into~\eqref{eq:microdynlaws}, for all $i,\,h\in \mathcal{I}$ we let
\begin{subequations}
\begin{empheq}[]{align}
	& (V_{i,t}'|I_t,\,{\bf V}_{t}) \sim T(v'_i|I_t,{\bf V}_{t}),\label{eq:intra_switcha}
  \\[5pt]
	& T(v'_i|h,\bv):= \begin{cases}
		K_i(v_i'|v_i) & \hbox{ if } i=h,\\
		\delta_{v_i}(v_i') & \hbox{ if } i\neq h,
	\end{cases}\label{eq:intra_switchb}
\end{empheq}
\end{subequations}
where $\delta_{(\cdot)}$ denotes the Dirac delta centred at $(\cdot)$. Moreover, recalling the notation~\eqref{eq:hats}, to incorporate (${\rm IIa}$) into~\eqref{eq:microdynlaws}, for all $i,i' \in \mathcal{I}$ we let
\begin{subequations}
\begin{empheq}[]{align}
	& (I'_t,\,{\bf V}_{t}''|I_t,\,{\bf V}_{t}) \sim \mathcal{T}(i',\bv''|I_t,{\bf V}_{t}),\label{eq:inter_switch_sponta}
  \\[5pt]
	& \mathcal{T}(i',\bv''|i, \bv):=
		\delta_{^{i'}\hat{\bv}}(^{i'}\hat{\bv}'')p(i'|i,v_i)P(v_{i'}''|i',i,v_i),\label{eq:inter_switch_spontb}
\end{empheq}
\end{subequations}
while to incorporate (${\rm IIb}$) for all $i,i'',k \in \mathcal{I}$ we let
\begin{subequations}
\begin{empheq}[]{align}
	& (I''_t,\,{\bf V}_{t}''|I_t,\,{\bf V}_{t},\,I^\ast_t,\,{\bf V}^\ast_{t}) \sim \mathcal{T}_k(i'',\bv''|I_t,{\bf V}_{t},I^\ast_t,{\bf V}^\ast_{t}),\label{eq:inter_switch_meda}
  \\[5pt]
	& \mathcal{T}_k(i'',\bv''|i,\bv,i^\ast,\bv^\ast):=		\delta_{^{i''}\hat{\bv}}(^{i''}\hat{\bv}'')q(i''|i,v_i,k,v_k^\ast)Q(v_{i''}''|i'',i,v_i,k,v_k^\ast).\label{eq:inter_switch_medb}
\end{empheq}
\end{subequations}
In summary, the definition~\eqref{eq:intra_switchb} translates in mathematical terms the idea that only individuals in compartment $i$ undergo changes in the trait represented by the structuring variable $v_i$ due to structuring-variable switching, according to the kernel $K_i$. Moreover, the definition~\eqref{eq:inter_switch_spontb} (resp. the definition~\eqref{eq:inter_switch_medb}) captures the fact that spontaneous compartment switching (resp. interaction-driven compartment switching) leads individuals in compartment $i$ with expression level $v_i$ of the trait represented by the structuring variable to transition into compartment $i'$ (resp. compartment $i''$), with probability $p(i'|i,v_i)$ (resp. with probability $q(i''|i,v_i,k,v_k^\ast)$, where $k$ and $v_k^\ast$ are the compartment and the value of the structuring variable of the individual with which/whom the interaction occurs), and acquire the expression level $v_{i'}''$ (resp. $v_{i''}''$) of the trait represented by the structuring variable of the new compartment, according to the kernel $P(v_{i'}''|i',i,v_i)$ (resp. the kernel $Q(v_{i''}''|i'',i,v_i,k,v_k^\ast)$).

\section{Mesoscopic model}\label{sec:meso}
Starting from the system~\eqref{eq:microdynlaws}, we now formally derive the mesoscopic model corresponding to the microscopic model presented in the previous section, which comprises a system of balance equations for the population density functions 
\beq
\label{eq:n}
n_i(t,v_i) \equiv n(t,i, v_i)  := \int_{^i\hat{\mathcal{V}}} \, f(t,i, \bv) \di ^i{\hat{\bv}}, \quad i \in \mathcal{I},
\eeq
where $f(t,i, \bv)$ is the probability density function given in~\eqref{eq:f}, and $^i\hat{\mathcal{V}}$ and $^i{\hat{\bv}}$ are defined via~\eqref{eq:hats}. The population density function $n_i(t,v_i)$ represents the distribution of individuals in compartment $i$ over the corresponding structuring-variable domain $\mathcal{V}_i$ at time $t$ (cf. the schematic in Figure~\ref{fig:meso}). Specifically, we will formally show that the dynamics of the population density functions $n_i(t,v_i)$ with $i \in \mathcal{I}$ are governed by the following IDE system
\begin{align}
    \resizebox{.93\textwidth}{!}{$\displaystyle
    \begin{aligned}[b]
        \partial_t n_i(t,v_i) &= \theta_i\Big[\int_{\mathcal{V}_i}K_i(v_i|v_i')n_i(t,v_i')\di v_i'-n_i(t,v_i)\Big] \\
        &\phantom{=} +\zeta\,r\,\Big[\sum_{\substack{j\in\mathcal{I} \\ j\neq i}}\int_{\mathcal{V}_j}p(i|j,v_j)P(v_{i}|i,j,v_j)
            n_j(t,v_j)\di v_j-\sum_{\substack{j\in\mathcal{I} \\ j\neq i}}p(j|i,v_i)\;n_i(t,v_i)\Big] \\
        &\phantom{=} +\zeta\,(1-r)\,\Big[\sum_{k\in\mathcal{I}}\sum_{\substack{j\in\mathcal{I} \\ j\neq i}}\int_{\mathcal{V}_j} \int_{\mathcal{V}_k}q(i|j,v_j,k,v_k)Q(v_i|i,j,v_j,k,v_k)n_k(t,v_k)n_j(t,v_j)\di v_k\di v_j \\
        &\phantom{=\zeta\,(1-r)\,\Big[}\;-\sum_{k\in\mathcal{I}}\sum_{\substack{j\in\mathcal{I} \\ j\neq i}}
            \int_{\mathcal{V}_k}q(j|i,v_i,k,v_k)n_k(t,v_k)\di v_k\;n_i(t,v_i)\Big],
                \quad v_i\in\mathcal{V}_i, \quad i\in\mathcal{I}.
    \end{aligned}
    $}
    \label{eq:IDE_i}
\end{align}

\paragraph{General evolution equation for expectations of observables.} We start by noting that, since the components of the random vector $\left(I_{t + \Delta t},\,{\bf V}_{t + \Delta t}\right)$ are given by~\eqref{eq:microdynlaws}, for any observable $\Phi : \mathcal{I} \times \mathcal{V} \to \mathbb{R}$ the expectation 
$$ \left<\Phi\left(I_{t},{\bf V}_{t}\right)\right>:=
    \sum_{l\in\mathcal{I}}\int_{\mathcal{V}}\Phi(l,\bv)f(t,l,\bv)\di\bv $$
satisfies (see, for instance,~\cite{pareschi2013BOOK})
\beq\label{eq:derivstep1}
\left<\Phi\left(I_{t + \Delta t}, {\bf V}_{t + \Delta t}\right) \right> = \left<\Phi\left(I_t, {\bf W}_{t}\right)\right> \, (1 - \zeta ) \, \Delta t + r\left<\Phi\left(I'_t, {\bf V}''_{t}\right)\right> \, \zeta  \, \Delta t + (1-r)\left<\Phi\left(I''_t, {\bf V}''_{t}\right)\right> \, \zeta  \, \Delta t ,
\eeq
where 
$$
{\bf W}_{t} \equiv \left(W_{1, t},\,\ldots,\,W_{N, t} \right) \; \text{ with } \; W_{i, t} := (1 - \Theta_i) \, V_{i, t} + \Theta_i \, V'_{i, t} \; \text{ for } \; i=1, \ldots, N
$$
and ${\bf V}''_{t} \equiv \left(V''_{1, t},\,\ldots,\,V''_{N, t} \right)$. Recalling that $Z$ and $\Theta_1,\,\ldots,\,\Theta_N$ are independent Bernoulli random variables with parameters $\zeta \Delta t$ and $\theta_1 \Delta t, \ldots, \theta_N \Delta t$, respectively,
introducing the diagonal matrix 
$$
{\bf \Theta} := {\rm diag}\left(\Theta_1,\,\ldots,\,\Theta_N \right)
$$
and rewriting the vector ${\bf W}_{t}$ as
$$
{\bf W}_{t} = \left({\bf I} -  {\bf \Theta} \right) \, {\bf V}_{t} + {\bf \Theta} \, {\bf V}'_{t},
$$
where ${\bf I}$ is the identity matrix, yields
\begin{equation}
    \begin{aligned}[b]
        \left<\Phi\left(I_t,{\bf W}_{t}\right)\right> &= \left<\Phi\left(I_t,{\bf V}_{t}\right)\right>\,{\rm P}({\bf\Theta}
            ={\bf 0})+\sum_{h\in\mathcal{I}}\left<\Phi\left(I_t,{\bf V}'^{,h}_{t}\right)\right>\,
                {\rm P}(\Theta_h =1\;\wedge\;\Theta_j=0\,\forall j\neq h) \\
        &\phantom{=} +\;o.t.\,,
    \end{aligned}
    \label{eq:derivstep1a}
\end{equation}
In~\eqref{eq:derivstep1a}, ${\rm P}(\cdot)$ is the notation that we will be using for probability, ${\bf 0}$ is the null matrix, 
\beq \label{eq:Vprime}
{\bf V}'^{,h}_{t} := \left(V_{1,t},\,\ldots, V_{h-1,t},\,V'_{h,t},\,V_{h+1,t},\,\ldots,\,V_{N,t}\right),
\eeq
and $o.t.$ are all the other terms accounting for the remaining cases in which multiple diagonal elements of the matrix ${\bf \Theta}$ are equal to $1$\red{, i.e. higher order terms in $\Delta t$ which will formally vanish in the limit $\Delta t \to 0$}. Using~\eqref{eq:derivstep1a}, we rewrite \eqref{eq:derivstep1} as
\begin{equation}
    \begin{aligned}[b]
        \left<\Phi\left(I_{t+\Delta{t}},{\bf V}_{t+\Delta{t}}\right)\right> &=
            \Big[\left<\Phi\left(I_t,{\bf V}_{t}\right)\right>\,{\rm P}({\bf\Theta}={\bf 0}) \\
        &\phantom{=} +\sum_{h\in\mathcal{I}}\left<\Phi\left(I_t,{\bf V}'^{,h}_{t}\right)\right>\,
            {\rm P}(\Theta_h=1\wedge\Theta_j=0\,\forall j\neq h)+o.t.\Big]\,(1-\zeta\Delta{t}) \\
        &\phantom{=} +\;r\,\left<\Phi\left(I'_t,{\bf V}''_{t}\right)\right>\,\zeta\Delta{t}+
            (1-r)\left<\Phi\left(I''_t,{\bf V}''_{t}\right)\right>\,\zeta\,\Delta{t}
    \end{aligned}
    \label{eq:derivstep2}
\end{equation}
and then, using the fact that, as proved in Appendix~\ref{appendix:A}, 
\beq \label{eq:asyres1}
 {\rm P}({\bf \Theta} = {\bf 0}) = \prod_{h \in \mathcal{I}} (1 - \theta_h \Delta t ) = 1 - \Delta t \sum_{h \in \mathcal{I}} \theta_h + o(\Delta t),
\eeq
$$ {\rm P}(\Theta_h=1\wedge\Theta_j=0\,\forall j\neq h)=\theta_h\Delta{t}
    \prod_{\substack{j\in\mathcal{I}\\ j\neq h}}(1-\theta_j\Delta{t})=\theta_h\Delta{t}+o(\Delta{t}) $$
and
\beq \label{eq:asyres3}
o.t. = o(\Delta t),
\eeq
with a little algebra we rewrite~\eqref{eq:derivstep2} as
\begin{equation}
    \begin{aligned}[b]
        \left<\Phi\left(I_{t+\Delta{t}},{\bf V}_{t+\Delta{t}}\right)\right> &=
            \left<\Phi\left(I_t,{\bf V}_{t}\right)\right>+\Delta{t}\sum_{h\in\mathcal{I}}\theta_h\Big(\left<\Phi\left(I_t,{\bf V}'^{,h}_{t}\right)\right>
                -\left<\Phi\left(I_t,{\bf V}_{t}\right)\right>\Big) \\
        &\phantom{=} +\;\Delta{t}\,\zeta\Big(r\left<\Phi\left(I'_t,{\bf V}''_{t}\right)\right>
            +(1-r)\left<\Phi\left(I''_t,{\bf V}''_{t}\right)\right>-\left<\Phi\left(I_t,{\bf V}_{t}\right)\right>\Big) \\
        &\phantom{=} +\;o(\Delta{t}).
    \end{aligned}
    \label{eq:derivstep3}
\end{equation}
From~\eqref{eq:derivstep3}, rearranging terms, dividing through by $\Delta t$ and letting $\Delta t \to 0^+$, we formally obtain the following evolution equation
\begin{equation}
    \begin{aligned}[b]
        \ddt\left<\Phi\left(I_{t},{\bf V}_t\right)\right> &= \sum_{h\in\mathcal{I}}\theta_h
            \Big(\left<\Phi\left(I_t,{\bf V}'^{,h}_{t}\right)\right>-\left<\Phi\left(I_t, {\bf V}_{t}\right)\right>\Big) \\
        &\phantom{=} +\,\zeta\Big(r\left<\Phi\left(I'_t, {\bf V}''_{t}\right)\right>
            +(1-r)\left<\Phi\left(I''_t,{\bf V}''_{t}\right)\right>-\left<\Phi\left(I_t,{\bf V}_{t}\right)\right>\Big),
    \end{aligned}
    \label{eq:EVPhi}
\end{equation}
which can be regarded as a weak form of the conservation equation for the probability density function $f(t,i,\bv)$. Expressing the expectations $\left< \cdot \right>$ in~\eqref{eq:EVPhi} in terms of sums and integrals against the probability density function $f$, and using \eqref{eq:intra_switcha}, \eqref{eq:inter_switch_sponta},  and \eqref{eq:inter_switch_meda}, the evolution equation~\eqref{eq:EVPhi} can then be rewritten as
\begin{align}
    \resizebox{.93\textwidth}{!}{$\displaystyle
    \begin{aligned}[b]
        \ddt &\sum_{l\in\mathcal{I}}\int_{\mathcal{V}}\Phi(l,\bv)\,f(t,l,\bv)\di\bv= \\
        &= \underbrace{\sum_{h\in\mathcal{I}}\theta_h\left[\sum_{l\in\mathcal{I}}\int_{\mathcal{V}}\Phi(l,\bv'^{,h})
            \left(\int_{\mathcal{V}_h}T(v_h'|l,\bv)\,f(t,l,\bv)\di v_h\right)\di\bv'^{,h}-
                \sum_{l\in\mathcal{I}}\int_{\mathcal{V}}\Phi(l,\bv)\,f(t,l,\bv)\di\bv\right]}_{=:\tcircle{i}} \\
        &\phantom{=} +\zeta\,r\underbrace{\sum_{l\in\mathcal{I}}\int_{\mathcal{V}}\Phi(l,\bv'')\left(\sum_{j\in\mathcal{I}}\int_{\mathcal{V}}
            \mathcal{T}(l,\bv''|j,\bv)f(t,j,\bv)\di\bv\right)\di\bv''}_{=:\tcircle{ii}} \\ 
        &\phantom{=} +\zeta\,(1-r)\underbrace{\sum_{l\in\mathcal{I}}\int_{\mathcal{V}}\Phi(l,\bv'')
            \left(\sum_{k\in\mathcal{I}}\sum_{j\in\mathcal{I}}\int_{\mathcal{V}}\int_{\mathcal{V}}\mathcal{T}_k(l,\bv''|j,\bv,k,v^\ast_k)
                f(t,j,\bv)f(t,k,\bv^\ast)\di\bv\di\bv^\ast\right)\di\bv''}_{=:\tcircle{iii}} \\
        &\phantom{=} -\zeta\,\underbrace{\sum_{l\in\mathcal{I}}\int_{\mathcal{V}}\Phi(l,\bv)\,f(t,l,\bv)\di\bv}_{=:\tcircle{iv}},
    \end{aligned}
    $}
    \label{eq:generic_form}
\end{align}
where, recalling the notation~\eqref{eq:Vprime}, $\bv'^{,h}=(v_1,\,\dots,\,v_{h-1},\,v_h',\,v_{h+1},\,\dots,\,v_N)$.

From~\eqref{eq:generic_form} we now derive a weak form of the IDE system~\eqref{eq:IDE_i}. To do so, we choose
\beq \label{eq:test_fun}
\Phi(l, \bv):=\delta_{i,l} \; \phi(v_i) \; \mathbbm{1}_{^i\hat{\mathcal{V}}}(^i\hat{\bv}),
\eeq
where $^i\hat{\mathcal{V}}$ and $^i\hat{\bv}$ are defined via~\eqref{eq:hats}, $\delta$ is the Kronecker delta, $\mathbbm{1}_{^i\hat{\mathcal{V}}}$ is the indicator function of the set $^i\hat{\mathcal{V}}$, and $\phi \in C^\infty(\mathcal{V}_i)$ is a smooth test function. We then substitute~\eqref{eq:test_fun} into~\eqref{eq:generic_form} and, for ease of presentation, carry out calculations for the left-hand side (LHS) of~\eqref{eq:generic_form} and the terms \tcircle{i}-\tcircle{iv} on the right-hand side (RHS) of~\eqref{eq:generic_form} one by one. 

\paragraph{\bf LHS of~\texorpdfstring{\eqref{eq:generic_form}}{}}
Recalling~\eqref{eq:n}, substituting~\eqref{eq:test_fun} into the LHS of~\eqref{eq:generic_form} yields
\beq\label{eq:lhsterm}
\ddt \sum_{l \in \mathcal{I}} \int_{\mathcal{V}} \Phi(l,\bv) \, f(t, l,\bv) \di \bv  = \int_{\mathcal{V}_i} \phi(v_i) \; \partial_t n_i(t,v_i) \,\di v_i.
\eeq

\paragraph{\tcircle{i} on the RHS of~\texorpdfstring{\eqref{eq:generic_form}}{}}
Recalling~\eqref{eq:n}, substituting~\eqref{eq:test_fun} along with~\eqref{eq:intra_switchb} into \tcircle{i} yields 
\begin{align*}
    \tcircle{i} &= \theta_i\left[\int_{\mathcal{V}_i}\int_{\mathcal{V}_i}\phi(v_i')K_i(v_i'|v_i)n_i(t,v_i)\di v_i'\di v_i
        -\int_{\mathcal{V}_i}\phi(v_i)n_i(t,v_i)\di v_i\right] \\
    &\phantom{=} +\sum_{\substack{h\in\mathcal{I} \\ h\neq i}}\theta_h\left[\int_{\mathcal{V}_i}\int_{\mathcal{V}_h}\phi(v_i)
        \delta_{v_h}(v_h')\,n_i(t,v_i)\di v'_h\di v_i-\int_{\mathcal{V}_i}\phi(v_i)n_i(t,v_i)\di v_i\right] \\
    &= \theta_i\left[\int_{\mathcal{V}_i}\int_{\mathcal{V}_i}\phi(v_i')K_i(v_i'|v_i)n_i(t,v_i)\di v_i'\di v_i
        -\int_{\mathcal{V}_i}\phi(v_i)n_i(t,v_i)\di v_i\right],
\end{align*}
from which, swapping $v_i'$ and $v_i$ in the first integral and then rearranging terms, we find
\begin{equation}\label{eq:circiterm} \tcircle{i}=\int_{\mathcal{V}_i}\phi(v_i)\left\{\theta_i\left[\int_{\mathcal{V}_i}K_i(v_i|v_i')n_i(t,v_i')\di v_i'
    -n_i(t,v_i)\right]\right\}\,\di v_i. \end{equation}

\paragraph{\tcircle{ii} on the RHS of~\texorpdfstring{\eqref{eq:generic_form}}{}}
Substituting first~\eqref{eq:test_fun} and then~\eqref{eq:inter_switch_spontb} into \tcircle{ii} and rearranging terms yields
\begin{align*}
    \tcircle{ii} &= \int_{\mathcal{V}_i}\phi(v_i'')\left(\sum_{j\in\mathcal{I}}\int_{^i\hat{\mathcal{V}}}\int_{\mathcal{V}}
        \mathcal{T}(i,\bv''|j,\bv)f(t,j,\bv)\di\bv\di ^i\hat{\bv}''\right)\di v_i'' \\
    &= \int_{\mathcal{V}_i}\phi(v_i'')\left(\sum_{\substack{j\in\mathcal{I} \\ j\neq i}}\int_{^i\hat{\mathcal{V}}}
        \int_{\mathcal{V}}\delta_{^{i}\hat{\bv}''}(^{i}\hat{\bv})p(i|j,v_j)P(v_i''|i,j,v_j)f(t,j,\bv)\di\bv\di ^i\hat{\bv}''\right)\di v_i'' \\
    &\phantom{=} +\int_{\mathcal{V}_i}\phi(v_i'')\left(\int_{^i\hat{\mathcal{V}}}\int_{\mathcal{V}}\delta_{^{i}\hat{\bv}''}(^{i}\hat{\bv})
        p(i|i,v_i)\delta_{v_i''}(v_i)f(t,i,\bv)\di\bv\di ^i\hat{\bv}''\right)\di v_i'' \\
    &= \int_{\mathcal{V}_i}\phi(v_i'')\left(\sum_{\substack{j\in\mathcal{I} \\ j\neq i}}\int_{\mathcal{V}}
        p(i|j,v_j)P(v_i''|i,j,v_j)f(t,j,\bv)\di\bv\right)\di v_i'' \\
    &\phantom{=} +\int_{\mathcal{V}_i}\phi(v_i'')\left(\int_{\mathcal{V}}p(i|i,v_i)\delta_{v_i}(v_i'')f(t,i,\bv)\di\bv\right)\di v_i'',
\end{align*}
from which, recalling \eqref{eq:n}, using the relation~\eqref{eq:ker_assump6a} and renaming $v_i''$ to $v_i$, we find
\begin{equation}\label{eq:circiiterm}   \tcircle{ii}=\int_{\mathcal{V}_i}\phi(v_i)
        \left\{\sum_{\substack{j\in\mathcal{I} \\ j\neq i}}\int_{\mathcal{V}_j}p(i|j,v_j)P(v_i|i,j,v_j)n_j(t,v_j)\di v_j
            +\left(1-\sum_{\substack{j\in\mathcal{I} \\ j\neq i}}p(j|i,v_i)\right)n_i(t,v_i)\right\}\di v_i. \end{equation}

\paragraph{\tcircle{iii} on the RHS of~\texorpdfstring{\eqref{eq:generic_form}}{}}
Substituting first~\eqref{eq:test_fun} and then~\eqref{eq:inter_switch_medb} into \tcircle{iii} and rearranging terms yields
\begin{align*}
    \resizebox{\textwidth}{!}{$\displaystyle
    \begin{aligned}
        \tcircle{iii} &= \int_{\mathcal{V}_i}\phi(v_i'')\left(\sum_{k\in\mathcal{I}}\sum_{j\in\mathcal{I}}
            \int_{^i\hat{\mathcal{V}}}\int_{\mathcal{V}}\int_{\mathcal{V}}\mathcal{T}_k(i,\bv''|j,\bv,k,v^\ast_k)f(t,j,\bv)f(t,k,\bv^\ast)
                \di\bv\di\bv^\ast\di ^i\hat{\bv}''\right)\di v_i'' \\
        &= \int_{\mathcal{V}_i}\phi(v_i'')\left(\sum_{k\in\mathcal{I}}\sum_{\substack{j\in\mathcal{I} \\ j\neq i}}
            \int_{^i\hat{\mathcal{V}}}\int_{\mathcal{V}}\int_{\mathcal{V}}\delta_{^{i}\hat{\bv}''}(^{i}\hat{\bv})
                q(i|j,v_j,k,v_k^\ast)Q(v_i''|i,j,v_j,k,v_k^\ast)f(t,j,\bv)f(t,k,\bv^\ast)\di\bv\di\bv^\ast\di ^i\hat{\bv}'' \right)\di v_i'' \\
        &\phantom{=} +\int_{\mathcal{V}_i}\phi(v_i'')\left(\sum_{k\in\mathcal{I}}\int_{^i\hat{\mathcal{V}}}\int_{\mathcal{V}}\int_{\mathcal{V}}
            \delta_{^{i}\hat{\bv}''}(^{i}\hat{\bv})q(i|i,v_i,k,v_k^\ast)\delta_{v_i''}(v_i)f(t,i,\bv)f(t,k,\bv^\ast)\di\bv\di\bv^\ast\di ^i\hat{\bv}''
                \right)\di v_i'' \\
        &= \int_{\mathcal{V}_i}\phi(v_i'')\left(\sum_{k\in\mathcal{I}}\sum_{\substack{j\in\mathcal{I} \\ j\neq i}}\int_{\mathcal{V}}
            \int_{\mathcal{V}}q(i|j,v_j,k,v_k^\ast)Q(v_i''|i,j,v_j,k,v_k^\ast)f(t,j,\bv)f(t,k,\bv^\ast)\di\bv\di\bv^\ast\right)\di v_i'' \\
        &\phantom{=} +\int_{\mathcal{V}_i}\phi(v_i'')\left(\sum_{k\in\mathcal{I}}\int_{\mathcal{V}}\int_{\mathcal{V}}q(i|i,v_i,k,v_k^\ast)
            \delta_{v_i}(v_i'')f(t,i,\bv)f(t,k,\bv^\ast)\di\bv\di\bv^\ast\right)\di v_i''.
    \end{aligned}
    $}
\end{align*}
Then, recalling \eqref{eq:n}, computing the integrals with respect to all the components of $\bv$ except $v_j$ and all the components of $\bv^\ast$ except $v_k^\ast$, using the relation~\eqref{eq:ker_assump6b} along with the fact that (cf. the integral identity~\eqref{ass:f})
$$
\sum_{k \in \mathcal{I}}\int_{\mathcal{V}_k}n_k(t,v_k^\ast) \di v_k^\ast =1  \;\; \forall \, t \in [0,\,\infty),
$$
and renaming $v_i''$ to $v_i$ and $v_k^\ast$ to $v_k$, we find
\begin{eqnarray}\label{eq:circiiiterm} 
    \tcircle{iii} &= \int_{\mathcal{V}_i}\phi(v_i)\left\{\sum_{k\in\mathcal{I}}
        \sum_{\substack{j\in\mathcal{I} \\ j\neq i}}\int_{\mathcal{V}_k}\int_{\mathcal{V}_j}q(i|j,v_j,k,v_k)Q(v_i|i,j,v_j,k,v_k)
            n_j(t,v_j)n_k(t,v_k)\di v_j\di v_k\right\}\di v_i \nonumber \\
    &\phantom{=} +\int_{\mathcal{V}_i}\phi(v_i)\left\{\left(1-\sum_{k\in\mathcal{I}}
        \sum_{\substack{j\in\mathcal{I} \\ j\neq i}}\int_{\mathcal{V}_k}q(j|i,v_i,k,v_k)n_k(t,v_k)\di v_k\right)n_i(t,v_i)\right\}\di v_i.
\end{eqnarray}

\paragraph{\tcircle{iv} on the RHS of~\texorpdfstring{\eqref{eq:generic_form}}{}}
Recalling~\eqref{eq:n}, substituting~\eqref{eq:test_fun} into \tcircle{iv} yields
\beq\label{eq:circivterm}
\tcircle{iv} = \int_{\mathcal{V}_i} \phi(v_i) \,  n_i(t,v_i) \, \di v_i.
\eeq

Substituting~\eqref{eq:lhsterm}, \eqref{eq:circiterm}, \eqref{eq:circiiterm}, \eqref{eq:circiiiterm}, and~\eqref{eq:circivterm} into~\eqref{eq:generic_form}, after a little algebra one obtains a weak formulation of the IDE system~\eqref{eq:IDE_i}.

\section{Macroscopic model}
\label{sec:macro}
In this section, we aim to recover a macroscopic representation of heterogeneously structured compartmental epidemiological systems starting from the mesoscopic representation provided by the IDE system~\eqref{eq:IDE_i}, i.e. to obtain a set of equations describing the time evolution of macroscopic quantities such as the fractions of individuals in the various compartments
$$ N_i(t):=\int_{\mathcal{V}_i}n_i(t,v_i)\di v_i, \qquad i\in\mathcal{I}, $$
and the mean values of the compartment-specific structuring variables
$$ M_i(t):=\frac{1}{N_i(t)}\int_{\mathcal{V}_i}v_in_i(t,v_i)\di v_i, \qquad i\in\mathcal{I}. $$

Integrating directly the IDE system~\eqref{eq:IDE_i} does not produce closed equations for these quantities. For instance, in the case of the $N_i$'s, noticing that
\begin{equation}
    \int_{\mathcal{V}_i}\left(\int_{\mathcal{V}_i}K_i(v_i|v_i')n_i(t,v_i')\di v_i'-n_i(t,v_i)\right)\di v_i=0
    \label{eq:Ki.cons_mass}
\end{equation}
because of~\eqref{eq:Ki_int1}, one finds
\begin{align*}
    \begin{aligned}[b]
        \frac{\di N_i}{\di t} &= \zeta r\sum_{\substack{j\in\mathcal{I} \\ j\neq i}}\left(
        \int_{\mathcal{V}_i}\int_{\mathcal{V}_j}p(i|j,v_j)P(v_{i}|i,j,v_j)n_j(t,v_j)\di v_j\di v_i 
            -\int_{\mathcal{V}_i}p(j|i,v_i)n_i(t,v_i)\di v_i\right) \\
        &\phantom{=} +\zeta(1-r)\sum_{k\in\mathcal{I}}\sum_{\substack{j\in\mathcal{I} \\ j\neq i}}\left(
        \int_{\mathcal{V}_i}\int_{\mathcal{V}_j}\int_{\mathcal{V}_k}
            q(i|j,v_j,k,v_k)Q(v_i|i,j,v_j,k,v_k)n_k(t,v_k)n_j(t,v_j)\di v_k\di v_j\di v_i\right. \\
	&\phantom{=+(1-r)\zeta\sum_{k\in\mathcal{I}}\sum_{\substack{j\in\mathcal{I} \\ j\neq i}}\left(\right.}
            -\left.\int_{\mathcal{V}_i}\int_{\mathcal{V}_k}q(j|i,v_i,k,v_k)n_k(t,v_k)n_i(t,v_i)\di v_k\di v_i\right),
    \end{aligned}
\end{align*}
which still requires the knowledge of the population density functions. To circumvent this difficulty, we adopt a procedure reminiscent of the \textit{hydrodynamic limit}, which in statistical mechanics allows one to obtain evolution equations for the hydrodynamic parameters of a gas (such as, for instance, the bulk density, velocity, and energy) from an underlying mesoscopic representation.

We introduce a small scaling parameter $0<\varepsilon\ll 1$ and let $\zeta=\varepsilon$ in~\eqref{eq:IDE_i}. Since $\zeta$ is the rate at which compartment switching occurs, this amounts to assuming a \textit{quasi-invariant} regime of transitions across the compartments. In other words, the probability that individuals leave their compartments over time is small. To compensate for such a smallness, thereby enabling the observation of significant time trends, we scale simultaneously the time variable as $t\to t/\varepsilon$. Upon introducing the scaled population density functions
$$ n_i^\varepsilon(t,v_i):=n_i\!\left(\tfrac{t}{\varepsilon},v_i\right), \qquad i\in\mathcal{I}, $$
whence $\partial_tn_i^\varepsilon=\frac{1}{\varepsilon}\partial_tn_i$, we rewrite~\eqref{eq:IDE_i} as
\begin{align}
    \resizebox{.93\textwidth}{!}{$\displaystyle
    \begin{aligned}[b]
        \varepsilon\partial_tn_i^\varepsilon(t,v_i) &= \theta_i\left(\int_{\mathcal{V}_i}K_i(v_i|v_i')n_i^\varepsilon(t,v_i')\di v_i'
            -n_i^\varepsilon(t,v_i)\right) \\
        &\phantom{=} +\varepsilon r\sum_{\substack{j\in\mathcal{I} \\ j\neq i}}\left(\int_{\mathcal{V}_j}p(i|j,v_j)P(v_i|i,j,v_j)n_j^\varepsilon(t,v_j)\di v_j
            -p(j|i,v_i)n_i^\varepsilon(t,v_i)\right) \\
        &\phantom{=} +\varepsilon(1-r)\sum_{k\in\mathcal{I}}\sum_{\substack{j\in\mathcal{I} \\ j\neq i}}\left(\int_{\mathcal{V}_j}\int_{\mathcal{V}_k}
            q(i|j,v_j,k,v_k)Q(v_i|i,j,v_j,k,v_k)n_k^\varepsilon(t,v_k)n_j^\varepsilon(t,v_j)\di v_k\di v_j\right. \\
	&\phantom{=+\varepsilon(1-r)\sum_{k\in\mathcal{I}}\sum_{\substack{j\in\mathcal{I} \\ j\neq i}}\left(\right.}
            -\left.n_i^\varepsilon(t,v_i)\int_{\mathcal{V}_k}q(j|i,v_i,k,v_k)n_k^\varepsilon(t,v_k)\di v_k\right).
    \end{aligned}
    $}
    \label{eq:IDE_i_resc1}
\end{align}
Integrating both sides with respect to $v_i$ and recalling~\eqref{eq:Ki.cons_mass}, we obtain
\begin{align}
    \resizebox{.93\textwidth}{!}{$\displaystyle
    \begin{aligned}[b]
        \ddt\int_{\mathcal{V}_i}n_i^\varepsilon(t,v_i)\di v_i &= r\sum_{\substack{j\in\mathcal{I} \\ j\neq i}}\int_{\mathcal{V}_i}\left(\int_{\mathcal{V}_j}p(i|j,v_j)P(v_i|i,j,v_j)n_j^\varepsilon(t,v_j)\di v_j
            -p(j|i,v_i)n_i^\varepsilon(t,v_i)\right)\di v_i \\
        &\phantom{=} +(1-r)\sum_{k\in\mathcal{I}}\sum_{\substack{j\in\mathcal{I} \\ j\neq i}}\int_{\mathcal{V}_i}\left(\int_{\mathcal{V}_j}\int_{\mathcal{V}_k}
            q(i|j,v_j,k,v_k)Q(v_i|i,j,v_j,k,v_k)\right. \\
        &\phantom{=} \qquad\qquad\qquad\qquad\qquad\qquad\qquad \times n_k^\varepsilon(t,v_k)n_j^\varepsilon(t,v_j)\di v_k\di v_j \\
	&\phantom{=} \qquad\qquad\qquad\qquad -\left.n_i^\varepsilon(t,v_i)\int_{\mathcal{V}_k}
            q(j|i,v_i,k,v_k)n_k^\varepsilon(t,v_k)\di v_k\right)\di v_i
    \end{aligned}
    $}
    \label{eq:IDE_i_resc2}
\end{align}
for all $\varepsilon>0$. To proceed further, we assume that the $K_i$'s are conservative also on average, i.e.
\begin{equation}
    \int_{\mathcal{V}_i}v_iK_i(v_i|v_i')\di v_i=v_i' \qquad \forall\,v_i'\in\mathcal{V}_i, \quad \forall\, i \in \mathcal{I},
    \label{eq:eig_assump2}
\end{equation}
so that
$$ \int_{\mathcal{V}_i}v_i\left(\int_{\mathcal{V}_i}K_i(v_i|v_i')n_i^\varepsilon(t,v_i')\di v_i'-n_i^\varepsilon(t,v_i)\right)\di v_i=0. $$
Hence, multiplying~\eqref{eq:IDE_i_resc1} by $v_i$ and integrating with respect to $v_i$ itself we find
\begin{align}
    \resizebox{.93\textwidth}{!}{$\displaystyle
    \begin{aligned}[b]
        \ddt\int_{\mathcal{V}_i}v_in_i^\varepsilon(t,v_i)\di v_i &=
            r\sum_{\substack{j\in\mathcal{I} \\ j\neq i}}\left(\int_{\mathcal{V}_j}p(i|j,v_j)\bar{P}(i,j,v_j)n_j^\varepsilon(t,v_j)\di v_j
                -\int_{\mathcal{V}_i}p(j|i,v_i)v_in_i^\varepsilon(t,v_i)\di v_i\right) \\
        &\phantom{=} +(1-r)\sum_{k\in\mathcal{I}}\sum_{\substack{j\in\mathcal{I} \\ j\neq i}}\left(\int_{\mathcal{V}_j}\int_{\mathcal{V}_k}q(i|j,v_j,k,v_k)
            \bar{Q}(i,j,v_j,k,v_k)\right. \\
        &\phantom{=} \qquad\qquad\qquad\qquad\qquad\qquad\qquad \times n_k^\varepsilon(t,v_k)n_j^\varepsilon(t,v_j)\di v_k\di v_j \\
	&\phantom{=} \qquad\qquad\qquad\qquad -\left.\int_{\mathcal{V}_i}v_in_i^\varepsilon(t,v_i)\int_{\mathcal{V}_k}
            q(j|i,v_i,k,v_k) n_k^\varepsilon(t,v_k)\di v_k\di v_i\right)
    \end{aligned}
    $}
    \label{eq:IDE_i_resc3}
\end{align}
for all $\varepsilon>0$, where we have used the following definitions:
$$ \bar{P}(i,j,v_j):=\int_{\mathcal{V}_i}v_iP(v_i|i,j,v_j)\di v_i, \qquad
    \bar{Q}(i,j,v_j,k,v_k):=\int_{\mathcal{V}_i}v_iQ(v_i|i,j,v_j,k,v_k)\di v_i. $$

To discover a universal trend valid in the regime of small $\varepsilon$ we pass now to the limit $\varepsilon\to 0^+$. To do so, we assume that $n_i^\varepsilon$ converges to some $n_i^0$, $i\in\mathcal{I}$, which, owing to~\eqref{eq:IDE_i_resc1}, formally satisfies
$$ \int_{\mathcal{V}_i}K_i(v_i|v_i')n_i^0(t,v_i')\di v_i'-n_i^0(t,v_i)=0. $$
Therefore, $n_i^0$ is an eigenfunction of the integral operator $\psi\mapsto\int_{\mathcal{V}_i}K(v_i|v_i')\psi(t,v_i')\di v_i'$ associated with the eigenvalue $1$ and such that
$$ \int_{\mathcal{V}_i}n_i^0(t,v_i)\di v_i=N_i^0(t), \qquad \int_{\mathcal{V}_i}v_in_i^0(t,v_i)\di v_i=N_i^0(t)M_i^0(t), $$
$N_i^0$, $M_i^0$ being the limit values of the scaled fraction and mean compartment-specific structuring variable of the individuals in compartment $i\in\mathcal{I}$. Following~\cite{lorenzi2023modelling}, we let $n_i^0(t,v_i)=N_i^0(t)\psi_i(t,v_i)$, where $\psi_i$ fulfills
$$ \int_{\mathcal{V}_i}\psi_i(t,v_i)\di v_i=1, \qquad \int_{\mathcal{V}_i}v_i\psi_i(t,v_i)\di v_i=M_i^0(t). $$
Then, letting, without loss of generality, $\mathcal{V}_i=[\ubar{v}_i,\,\bar{v}_i]$ with $\ubar{v}_i<\bar{v}_i$, under assumptions~\eqref{eq:Ki_supp},~\eqref{eq:Ki_int1},~\eqref{eq:Ki_var}, and~\eqref{eq:eig_assump2}, it is possible to prove, see again~\cite{lorenzi2023modelling}, that there exists a unique $\psi_i$, which is in the form
$$ \psi_i(t,v_i)=(1-M_i^0(t))\delta_{\ubar{v}_i}(v_i)+M_i^0(t)\delta_{\bar{v}_i}(v_i),
    \qquad i\in\mathcal{I}, $$
so that finally there exists a unique $n_i^0$, which is in the form
$$ n_i^0(t,v_i)=N_i^0(t)(1-M_i^0(t))\delta_{\ubar{v}_i}(v_i)+N_i^0(t)M_i^0(t)\delta_{\bar{v}_i}(v_i),
    \qquad i\in\mathcal{I}. $$
In the following, without loss of generality, we set $\ubar{v}_i=0$ and $\bar{v}_i=1$, that is,
\begin{equation}
    \mathcal{V}_i=[0,\,1] \qquad \forall\,i\in\mathcal{I}.
    \label{eq:defVi}
\end{equation}

Substituting the expressions for the $n_i^0$'s just found into~\eqref{eq:IDE_i_resc2}, after passing there to the limit $\varepsilon\to 0^+$, and dropping the superscripts ``$0$'' for convenience, yields
\begin{align}
    \begin{aligned}[b]
        \frac{\di N_i}{\di t} &= r\sum_{\substack{j\in\mathcal{I} \\ j\neq i}}\Bigl[\Bigl(p(i|j,0)(1-M_j)+p(i|j,1)M_j\Bigr)N_j
            -\Bigl(p(j|i,0)(1-M_i)+p(j|i,1)M_i\Bigl)N_i\Bigr] \\
        &\phantom{=} +(1-r)\sum_{k\in\mathcal{I}}\sum_{\substack{j\in\mathcal{I} \\ j\neq i}}\Bigl[\Bigl(q(i|j,0,k,0)(1-M_j)(1-M_k)+
            q(i|j,1,k,0)M_j(1-M_k) \\
        &\phantom{=} \qquad\qquad\qquad\qquad +q(i|j,0,k,1)(1-M_j)M_k+q(i|j,1,k,1)M_jM_k\Bigr)N_j \\
        &\phantom{=} \qquad\qquad\qquad\qquad -\Bigl(q(j|i,0,k,0)(1-M_i)(1-M_k)+q(j|i,1,k,0)M_i(1-M_k) \\
        &\phantom{=} \qquad\qquad\qquad\qquad +q(j|i,0,k,1)(1-M_i)M_k+q(j|i,1,k,1)M_iM_k\Bigr)N_i\Bigr]N_k.
    \end{aligned}
    \label{eq:ODE_Ni_01}
\end{align}
Note that, as expected, the ODE system~\eqref{eq:ODE_Ni_01} is mass-preserving, that is,
\begin{equation}
\label{eq:masspresmacro}
\ddt\sum_{i\in\mathcal{I}}N_i(t) =0 \;\; \forall \, t>0 \quad \Longrightarrow \quad \sum_{i\in\mathcal{I}}N_i(t) = \sum_{i\in\mathcal{I}}N_i(0)  \;\; \forall \, t>0.
\end{equation}

Similarly, substituting the expressions for the $n_i^0$'s found above into~\eqref{eq:IDE_i_resc3}, after passing also there to the limit $\varepsilon\to 0^+$, and dropping again the superscripts ``$0$'', gives
\begin{align}
    \begin{aligned}[b]
        \ddt(N_iM_i) &= r\sum_{\substack{j\in\mathcal{I} \\ j\neq i}}\Bigl[\Bigl(p(i|j,0)\bar{P}(i,j,0)(1-M_j)+p(i|j,1)\bar{P}(i,j,1)M_j\Bigr)N_j
            -p(j|i,1)N_iM_i\Bigr] \\
        &\phantom{=} +(1-r)\sum_{k\in\mathcal{I}}\sum_{\substack{j\in\mathcal{I} \\ j\neq i}}\Bigl[\Bigl(q(i|j,0,k,0)\bar{Q}(i,j,0,k,0)(1-M_j)(1-M_k) \\
        &\phantom{=} \qquad\qquad\qquad\qquad +q(i|j,1,k,0)\bar{Q}(i,j,1,k,0)M_j(1-M_k) \\
        &\phantom{=} \qquad\qquad\qquad\qquad +q(i|j,0,k,1)\bar{Q}(i,j,0,k,1)(1-M_j)M_k \\
        &\phantom{=} \qquad\qquad\qquad\qquad +q(i|j,1,k,1)\bar{Q}(i,j,1,k,1)M_jM_k\Bigr)N_j \\
        &\phantom{=} \qquad\qquad\qquad\qquad -\Bigl(q(j|i,1,k,0)M_i(1-M_k)+q(j|i,1,k,1)M_iM_k\Bigr)N_i\Bigr]N_k.
    \end{aligned}
    \label{eq:ODE_NiMi_01}
\end{align}

The ODE system~\eqref{eq:ODE_Ni_01},\eqref{eq:ODE_NiMi_01} constitutes, for $i\in\mathcal{I}$, a self-consistent compartmental model describing the evolution in time of the macroscopic quantities $N_i$, $M_i$ in terms of the microscopic information contained in the model parameter $r$ and  functions $p$, $q$, $\bar{P}$, and $\bar{Q}$.
	
\subsection{The homogeneous case}
A considerable simplification of the ODE system~\eqref{eq:ODE_Ni_01},\eqref{eq:ODE_NiMi_01} is obtained in the \textit{homogeneous} case, i.e. when one assumes that the functions $p$, $q$, $\bar{P}$, and $\bar{Q}$ depend on the compartment indices but are independent of the structuring variables, that is, 
\begin{align*}
   & p(j|i,v_i)\equiv p(j|i),  & & \bar{P}(i,j,v_j)\equiv\bar{P}(i,j), \\ 
   & q(j|i,v_i,k,v_k)\equiv q(j|i,k), & & \bar{Q}(i,j,v_j,k,v_k)\equiv\bar{Q}(i,j,k).
\end{align*}
In this case, recalling~\eqref{eq:ker_assump5}, we then find that~\eqref{eq:ODE_Ni_01} can be written in the form
\begin{equation}
    \frac{\di N_i}{\di t}=r\sum_{\substack{j\in\mathcal{I} \\ j\neq i}}\bigl(p(i|j)N_j-p(j|i)N_i\bigr)
        +(1-r)\sum_{k\in\mathcal{I}}\sum_{\substack{j\in\mathcal{I} \\ j\neq i}}\bigl(q(i|j,k)N_j-q(j|i,k)N_i\bigr)N_k, \qquad i\in\mathcal{I}.
    \label{eq:ODE_i_hom}
\end{equation}

Notice that~\eqref{eq:ODE_i_hom} is a self-consistent system in the unknowns $N_i$, $i\in\mathcal{I}$. In other words, in the homogeneous case the evolution of the fractions of individuals is independent of that of the mean values of the compartment-specific structuring variables. The same is not true in the general case~\eqref{eq:ODE_Ni_01}, owing to the coupling produced by the inhomogeneous coefficients $p$ and $q$. This result makes it possible to assimilate~\eqref{eq:ODE_i_hom} to a classical compartmental model of population dynamics. On the other hand, unlike a classical compartmental model, the framework defined by~\eqref{eq:ODE_Ni_01},\eqref{eq:ODE_NiMi_01} still allows one to track the evolution in time of the mean values of the compartment-specific structuring variables in consequence of the transitions of the individuals between the various compartments. In fact, from~\eqref{eq:ODE_NiMi_01}, under the homogeneity assumption, we find
\begin{align*}
    \ddt(N_iM_i) &= r\left(\sum_{\substack{j\in\mathcal{I} \\ j\neq i}}p(i|j)\bar{P}(i,j)N_j-(1-p(i|i))N_iM_i\right) \\
    &\phantom{=} +(1-r)\sum_{k\in\mathcal{I}}\sum_{\substack{j\in\mathcal{I} \\ j\neq i}}\bigl(q(i|j,k)\bar{Q}(i,j,k)N_j
        -q(j|i,k)N_iM_i\bigr)N_k, \qquad i\in\mathcal{I},
\end{align*}
which, developing the time derivative on the left-hand side and using~\eqref{eq:ODE_i_hom}, further becomes
\begin{align}
    \begin{aligned}[b]
        \frac{\di M_i}{\di t} &= \frac{r}{N_i}\sum_{\substack{j\in\mathcal{I} \\ j\neq i}}p(i|j)\bigl(\bar{P}(i,j)-M_i\bigr)N_j \\
        &\phantom{=} +\frac{1-r}{N_i}\sum_{k\in\mathcal{I}}\sum_{\substack{j\in\mathcal{I} \\ j\neq i}}q(i|j,k)\bigl(\bar{Q}(i,j,k)-M_i\bigr)N_jN_k,
            \qquad i\in\mathcal{I}.
    \end{aligned}
    \label{eq:ODE_M_i_hom}
\end{align}

We notice that when $N_i$ approaches $0$ a singularity in~\eqref{eq:ODE_M_i_hom} has to be expected. Moreover, we observe that the right-hand side of~\eqref{eq:ODE_M_i_hom} is linear in $M_i$ and, therefore, the existence of an endemic equilibrium $(N_i^\ast)_{i\in\mathcal{I}}\in (0,\,1)^N$ of~\eqref{eq:ODE_i_hom} implies straightforwardly the existence of a corresponding equilibrium $(M_i^\ast)_{i\in\mathcal{I}}\in [0,\,1]^N$. In such a case, if $(N_i^\ast)_{i\in\mathcal{I}}$ is asymptotically stable then $(M_i^\ast)_{i\in\mathcal{I}}$ inherits the same stability property.

\subsubsection{Basic reproduction number~\texorpdfstring{$\boldsymbol{\mathcal{R}_0}$}{}}

In this section, we apply the Next Generation Matrix method \cite{diekmann1990definition,diekmann2010construction,van2002reproduction} to derive the basic reproduction number $\mathcal{R}_0$ of the ODE system \eqref{eq:ODE_i_hom}. We remark that the construction presented here reduces the method to its most elementary building blocks, which are the probabilities of leaving each infectious compartments and the probabilities of contaminating a susceptible individual upon interacting with them. This way, we elucidate the structural microscopic origin of an aggregate parameter as significant as $\mathcal{R}_0$ in the context of compartmental epidemiological models.

Without loss of generality, we assume that $i=1$ corresponds to the Susceptible compartment. We focus on systems of ODEs of type~\eqref{eq:ODE_i_hom} which possess only one Disease Free Equilibrium (DFE) of the form
\begin{equation}
    (N_1,\,N_2,\,\dots,\,N_N)=(1,\,0,\,\dots,\,0).
    \label{eq:hom_DFE}
\end{equation}
Such systems are quite common in the literature, see Section~\ref{sec:examples} for specific case studies.

The entry $(i,\,j)$ of the Jacobian matrix $\mathbf{J}$ of~\eqref{eq:ODE_i_hom} is given by
\begin{align*}
    J_{ij} &= r\bigl(p(i|j)-\delta_{i,j}\bigr) \\
    &\phantom{=} +(1-r)\left(\sum_{\substack{l\in\mathcal{I} \\ l\neq i}}\bigl(q(i|l,j)N_l-q(l|i,j)N_i\bigr)
        +\sum_{k\in\mathcal{I}}q(i|j,k)N_k(1-\delta_{i,j})
            -\sum_{k\in\mathcal{I}}\sum_{\substack{l\in\mathcal{I} \\ l\neq i}}q(l|i,k)N_k\delta_{i,j}\right),
\end{align*}
where $\delta$ is the Kronecker delta, cf.~\eqref{eq:test_fun}. Evaluating at the DFE~\eqref{eq:hom_DFE}, this produces
\begin{align*}
    J_{ij}(\text{DFE}) &= r\bigl(p(i|j)-\delta_{i,j}\bigr) \\
    &\phantom{=} +(1-r)\left(q(i|1,j)(1-\delta_{i,1})-\sum_{\substack{l\in\mathcal{I} \\ l\neq i}}q(l|i,j)\delta_{i,1}
        +q(i|j,1)(1-\delta_{i,j})-\sum_{\substack{l\in\mathcal{I} \\ l\neq i}}q(l|i,1)\delta_{i,j}\right).
\end{align*}

Next, we consider the sub-matrix $\mathbf{J}^\ast(\text{DFE})$ of $\mathbf{J}(\text{DFE})$ determined by infectious or infection-related (such as, for instance, Exposed, Asymptomatic, etc) compartments and we seek a decomposition of the form
$$ \mathbf{J}^\ast(\text{DFE})=\mathbf{A}-\mathbf{B}, $$
where the matrix $\mathbf{A}$ has non-negative entries and represents \emph{transmission}, whereas the matrix $\mathbf{B}$ is invertible and represents \emph{transitions}. There is not a unique way to define these two matrices. However, considering their respective interpretations, natural definitions for their entries are
\begin{align}
    \begin{aligned}[c]
        A_{ij} &:= (1-r)\left(q(i|1,j)(1-\delta_{i,1})-\sum_{\substack{l\in\mathcal{I} \\ l\neq i}}q(l|i,j)\delta_{i,1}
            +q(i|j,1)(1-\delta_{i,j})-\sum_{\substack{l\in\mathcal{I} \\ l\neq i}}q(l|i,1)\delta_{i,j}\right), \\
        B_{ij} &:= r\bigl(\delta_{i,j}-p(i|j)\bigr)
    \end{aligned}
    \label{eq:MandV}
\end{align}
for all $i,\,j\in\mathcal{I}$ associated with infectious or infection-related compartments. Notice that, owing to~\eqref{eq:ker_assump5c}, Gershgorin first theorem applied to the columns of $\mathbf{B}$ ensures that all the eigenvalues of $\mathbf{B}$ have strictly non-negative real parts. Nevertheless, the origin of the complex plane may be contained in each of the Gershgorin discs; hence, we do not know in general whether $\mathbf{B}$ is invertible or not. A simple criterion ensuring invertibility is that there exists at least one non-infectious compartment that can be reached with a spontaneous flow from each infectious compartment. This will be the case in all the specific models we shall present in Section~\ref{sec:examples}.

Assuming that $\mathbf{B}$ in~\eqref{eq:MandV} is invertible, the Next Generation Matrix method provides the basic reproduction number $\mathcal{R}_0$ as
\begin{equation}
    \mathcal{R}_0=\rho(\mathbf{A}\mathbf{B}^{-1}),
    \label{eq:Rzero}
\end{equation}
where $\rho$ denotes the spectral radius. 

One of the most classical results on compartmental models is the global stability of the DFE under the condition $\mathcal{R}_0<1$. In addition to this, the modelling framework based on~\eqref{eq:ODE_i_hom} allows one to build compartmental models which exhibit both forward and backward bifurcations at $\mathcal{R}_0=1$. In the former case, the DFE usually loses stability as $\mathcal{R}_0$ becomes greater than 1, in favour of the stability of a (unique) Endemic Equilibrium (EE). In the latter case, there exists a value $\mathcal{R}_0^\ast\in (0,\,1)$ such that for $\mathcal{R}_0^\ast<\mathcal{R}_0<1$ the system exhibits bi-stability of DFE and EE as well as the existence of an unstable EE at the boundary of the corresponding basins of attraction.

Another classical result is the existence of at least one EE when $\mathcal{R}_0>1$. However, the modelling framework based on~\eqref{eq:ODE_i_hom} presented here can easily be adapted to at least one system which does not possess this characteristic, namely the celebrated SIR model. Under the assumption $\mathcal{R}_0>1$, SIR orbits are indeed heteroclinic to the set of null infection (see e.g.~\cite[Lemma 1]{jardon2021geometric} for a proof) and the system does not possess any EE. Consequently, we cannot conclude, in general, the existence of any EE for system~\eqref{eq:ODE_i_hom} without referring to specific cases or enforcing additional assumptions on the model functions $p$, $q$, $\bar{P}$, and $\bar{Q}$.

\section{Case studies}
\label{sec:examples}
In this section, we demonstrate, through a few case studies, how different classical compartmental epidemiological models can be derived from the ODE system~\eqref{eq:ODE_i_hom}, under appropriate choices of the microscopic parameter functions $p$, $q$, $\bar{P}$, and $\bar{Q}$. In contrast to their classical counterparts, the models here derived comprise also a system of ODEs for the mean values of the corresponding compartment-specific structuring variables, which is obtained from the ODE system~\eqref{eq:ODE_M_i_hom}, thus providing a richer description of the dynamics of the epidemiological system considered. 

For every case study, we specify the choices of the microscopic parameter functions $p$, $q$, $\bar{P}$, and $\bar{Q}$ under which~\eqref{eq:ODE_i_hom} reduces to the ODE system of the classical model. 
Moreover, we compare numerical solutions and analytical results of the ODEs of the macroscopic model with the results of Monte Carlo simulations of the underlying individual-based model, and show that there is excellent agreement between them. This provides validation of the formal procedures employed in Sections~\ref{sec:meso} and \ref{sec:macro} to obtain first the IDE system~\eqref{eq:IDE_i} and then the ODE systems~\eqref{eq:ODE_i_hom} and~\eqref{eq:ODE_M_i_hom} from the individual-based model governed by the system~\eqref{eq:microdynlaws}. \red{Simulations were performed in Matlab (version 2023a). The macroscopic ODE systems were solved using standard numerical routines (i.e. \texttt{ode45} in MATLAB 2023a), as they are low-dimensional and not stiff. By contrast, due to large number of agents, Monte Carlo simulations of the individual-based model required more attention. In particular, algorithms were implemented avoiding unnecessary loops and making use of vectorised operations, which significantly improved efficiency. As a result, we were able to carry out simulations involving up to $10^6$ agents per run on a standard personal computer (12th Gen Intel(R) Core(TM) i7-1260P, 2.10 GHz, 12 cores, 16 logical processors, 16 GB RAM). We note that more complex scenarios may require larger numbers of agents to ensure statistical convergence, in which case additional computational resources would be beneficial.
}

For consistency with the extant literature, in each case study we use an alphabetic index set $\mathcal{I}$ rather than a numerical one. For instance, in the case of three compartments of Susceptible, Infectious, and Recovered individuals, we set $\mathcal{I}=\{S,\,I,\,R\}$ instead of $\mathcal{I}=\{1,\,2,\,3\}$. Moreover, for consistency with~\eqref{eq:defVi}, we choose $\mathcal{V}_i=[0,\,1]$ for all $i \in \mathcal{I}$. The values of key parameters of the macroscopic models used in numerical simulations are provided in the captions of Figures~
\ref{fig:SIRS_simu}, \ref{fig:SIRS2_simu}, \ref{fig:SIRWS_simu}, while the definitions of the model parameters and functions that are used to carry out Monte Carlo simulations of the individual-based models are provided in Appendix~\ref{sec:appibm}. Finally, the initial conditions for the fractions of individuals in the various compartments are such that, consistently with~\eqref{ass:f}, the following normalisation condition holds:
$$
\sum_{i\in\mathcal{I}} N_i(0) =1.
$$
Since the macroscopic model and the underlying microscopic model are mass-preserving, cf. \eqref{eq:masspresmacro}, such a normalisation condition then holds for all $t>0$, i.e.
$$
\sum_{i\in\mathcal{I}}N_i(t)=1 \;\; \forall \, t>0.    
$$

We notice that in all cases considered here the components of the unique DFE, cf. \eqref{eq:hom_DFE}, are $N_S=1$ and $N_j=0$ for all $j \in \mathcal{I}$ with $j \neq S$.

\subsection{SIRS model}
\label{sec:exSIRS}
\begin{figure}[!t]
\centering
\begin{tikzpicture}
    \node[draw,circle,thick,minimum size=.75cm] (s) at (0,0) {$N_S$};
    \node[draw,circle,thick,minimum size=.75cm] (i) at (4.5,0) {$N_I$};
    \node[draw,circle,thick,minimum size=.75cm] (r) at (4.5,-2) {$N_R$};
    \draw[-{Latex[length=2.mm, width=1.5mm]},thick] (s)--(i) node[above, midway]{\small $(1-r)q(I|S,I)N_SN_I$};
    \draw[-{Latex[length=2.mm, width=1.5mm]},thick] (i)--(r) node[right, midway]{\small $rp(R|I)N_I$};
    \draw[{Latex[length=2.mm, width=1.5mm]}-,thick] (s)--(r) node[below,midway]{\small $rp(S|R)N_R\qquad\qquad$};
\end{tikzpicture}
\caption{Flow diagram of the SIRS model}
\label{fig:SIRS}
\end{figure}
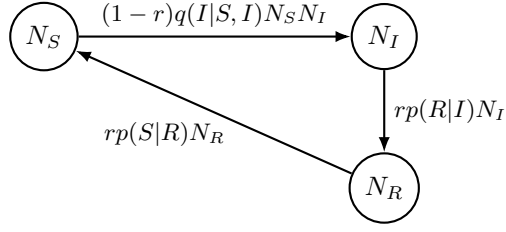

Considering an SIRS system, we take $N=3$ and $\mathcal{I}=\{S,\,I,\,R\}$. As per the diagram displayed in Figure~\ref{fig:SIRS}, only the following types of compartment switching occur: $S\to I$ (driven by the interaction between a susceptible and an infectious individual), $I\to R$ (spontaneous), and $R\to S$ (spontaneous). Therefore, the non-vanishing transition probabilities are $q(I|S,I)$, $p(R|I)$, and $p(S|R)$, whereas all the remaining $p$'s and $q$'s are zero. The ODE system~\eqref{eq:ODE_i_hom} then reduces to
\begin{equation}
    \begin{cases}
        \dfrac{\di N_S}{\di t}=rp(S|R)N_R-(1-r)q(I|S,I)N_SN_I, \\[3mm]
        \dfrac{\di N_I}{\di t}=-rp(R|I)N_I+(1-r)q(I|S,I)N_SN_I, \\[3mm]
        \dfrac{\di N_R}{\di t}=r\bigl(p(R|I)N_I-p(S|R)N_R\bigr).
    \end{cases}
    \label{eq:exSIRS}
\end{equation}
In particular, the matrices defined via~\eqref{eq:MandV} are scalar quantities, i.e.
$$\mathbf{A} \equiv A=(1-r)q(I|S,I), \qquad \mathbf{B} \equiv B=r(1-p(I|I))=rp(R|I), $$
where the above expression for $B$ follows from \eqref{eq:ker_assump5c}, recalling that in this case $p(S|I)=0$. Finally, through~\eqref{eq:Rzero}, we find
$$ \mathcal{R}_0=\frac{A}{B}=\frac{(1-r)q(I|S,I)}{rp(R|I)}. $$
This coincides with the basic reproduction number classically computed as the ratio between the rate at which susceptible individuals become infectious and that at which infectious individuals become recovered, which from~\eqref{eq:exSIRS} are indeed $(1-r)q(I|S,I)$ and $rp(R|I)$, respectively.

The DFE $(N_S,\,N_I,\,N_R)=(1,\,0,\,0)$ is globally asymptotically stable when $\mathcal{R}_0<1$, cf.~\cite{cangiotti2023survey}. Conversely, when $\mathcal{R}_0>1$ there is a unique EE of components
$$ N_S^\ast=\frac{1}{\mathcal{R}_0}, \qquad N_I^\ast=\frac{p(S|R)}{p(S|R)+p(R|I)}\left(1-\frac{1}{\mathcal{R}_0}\right),
    \qquad N_R^\ast=1-N_S^\ast-N_I^\ast, $$
which is globally stable, cf.~\cite{o2010lyapunov}.

Moreover, in this case the compartment-specific structuring variables can be interpreted as:
\begin{enumerate*}[label=(\roman*)]
    \item the level of \textit{resistance to infection} in compartment $i=S$;
    \item the level of \textit{viral load} in compartment $i=I$;
    \item the level of \textit{immunity} in compartment $i=R$.
\end{enumerate*}
The evolution of their mean values is governed by the following ODE system, which is derived from~\eqref{eq:ODE_M_i_hom}:
\begin{equation}
    \begin{cases}
	   \dfrac{\di M_S}{\di t}=r\dfrac{p(S|R)(\bar{P}(S,R)-M_S)}{N_S}N_R, \\[3mm]
	   \dfrac{\di M_I}{\di t}=(1-r)q(I|S,I)(\bar{Q}(I,S,I)-M_I)N_S, \\[3mm]
	   \dfrac{\di M_R}{\di t}=r\dfrac{p(R|I)(\bar{P}(R,I)-M_R)}{N_R}N_I,
    \end{cases}
    \label{eq:exSIRS_Mi}
\end{equation}
the equilibrium of which corresponding to the EE of~\eqref{eq:exSIRS} is 
$$ M_S^\ast=\bar{P}(S,R), \qquad M_I^\ast=\bar{Q}(I,S,I), \qquad M_R^\ast=\bar{P}(R,I). $$

\begin{figure}[!t]
\centering
\subfigure[]{\includegraphics[width=.45\textwidth]{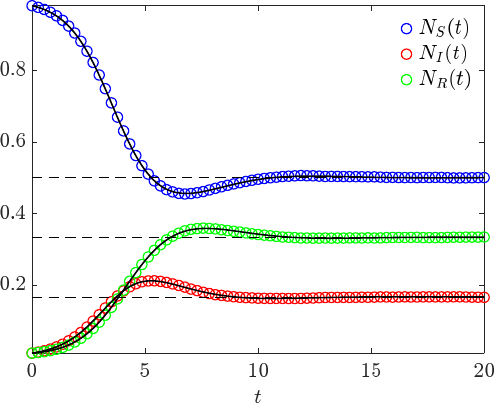}} \qquad
\subfigure[]{\includegraphics[width=.45\textwidth]{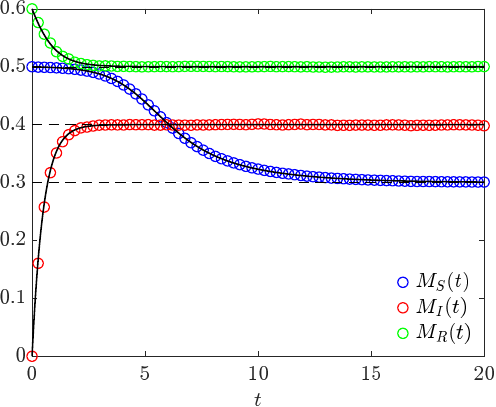}}
\caption{SIRS model. Solid lines: numerical solutions  of~\eqref{eq:exSIRS} (panel (a)) and~\eqref{eq:exSIRS_Mi} (panel (b)); circles: results of Monte Carlo simulations of the individual-based model governed by~\eqref{eq:microdynlaws} adapted to this case; dashed lines: analytical equilibria. Relevant parameters: 
$rp(S|R)=0.05$, $(1-r)q(I|S,I)=0.2$, $rp(R|I)=0.1$, $\bar{P}(S,R)=0.3$, $\bar{Q}(I,S,I)=0.4$, $\bar{P}(R,I)=0.5$. Hence, $\mathcal{R}_0=2$. Initial condition: $(N_S(0),\,N_I(0),\,N_R(0),\,M_S(0),\,M_I(0),\,M_R(0))=(0.98,\,0.01,\,0.01,\,0.5,\,0,\,0.6)$. Monte Carlo simulations are carried out using $10^6$ agents and defining the model parameters and functions as detailed in Appendix~\ref{sec:appibm1}
}
\label{fig:SIRS_simu}
\end{figure}

The plots in Figure~\ref{fig:SIRS_simu} show a comparison between numerical solutions of the ODE systems~\eqref{eq:exSIRS} and \eqref{eq:exSIRS_Mi} and the results of Monte Carlo simulations of the corresponding individual-based model for the case where  $\mathcal{R}_0>1$.

\subsection{SIRS model with secondary infections}
\begin{figure}[!t]
\centering
\begin{tikzpicture}
    \node[draw,circle,thick,minimum size=.75cm] (s) at (1,0) {$N_S$};
    \node[draw,circle,thick,minimum size=.75cm] (i) at (1,-2) {$N_I$};
    \node[draw,circle,thick,minimum size=.75cm] (t) at (5,-2) {$N_T$};
    \node[draw,circle,thick,minimum size=.75cm] (p) at (5,0) {$N_P$};
    \node[draw,circle,thick,minimum size=.75cm] (y) at (10,0) {$N_Y$};
    \node[draw,circle,thick,minimum size=.75cm] (r) at (10,-2) {$N_R$};
    \draw[-{Latex[length=2.mm, width=1.5mm]},thick] (s)--(i) node[above, midway]{};
    \draw[-{Latex[length=2.mm, width=1.5mm]},thick] (i)--(t) node[below, midway]{$rp(T|I)N_I$};
    \draw[-{Latex[length=2.mm, width=1.5mm]},thick] (t)--(p) node[left, midway]{$rp(P|T)N_T$};
    \draw[-{Latex[length=2.mm, width=1.5mm]},thick] (p)--(y) node[above, midway]{};
    \draw[-{Latex[length=2.mm, width=1.5mm]},thick] (p)--(s) node[above, midway]{$rp(S|P)N_P$};
    \draw[-{Latex[length=2.mm, width=1.5mm]},thick] (y)--(r) node[right, midway]{$rp(R|Y)N_Y$};
    \draw[-{Latex[length=2.mm, width=1.5mm]},thick] (r)--(p) node[below, midway]{$rp(P|R)N_R\quad\quad\quad$};
\end{tikzpicture}
\caption{Flow diagram of the SIRS model with secondary infections. The infection rates between $N_S$ and $N_I$ (primary) and between $N_P$ and $N_Y$ (secondary) are omitted to avoid overcrowding}
\label{fig:SIRSsquared}
\end{figure}
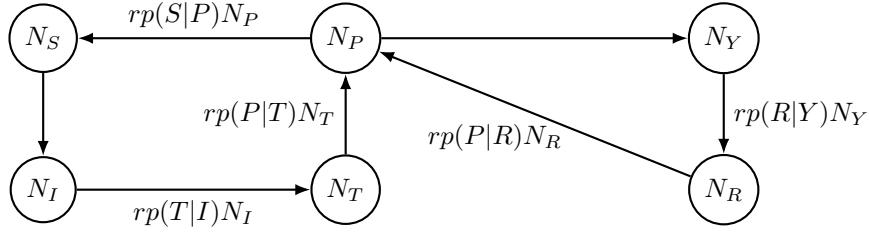

Within the general formulation provided by~\eqref{eq:ODE_i_hom}, we can model systems exhibiting either \emph{forward} bifurcations when $\mathcal{R}_0=1$, such as the SIRS model presented in Section~\ref{sec:exSIRS}, or \emph{backward} bifurcations. In the former case, one usually expects global asymptotic stability of the DFE when $\mathcal{R}_0<1$, and global asymptotic stability of the EE when $\mathcal{R}_0>1$. On the contrary, in the latter case there exists a value $\mathcal{R}_0^\ast\in (0,\,1)$ of the basic reproduction number such that for $\mathcal{R}_0^\ast<\mathcal{R}_0<1$ the system exhibits bi-stability between the DFE and one EE along with a second unstable EE. This happens, for instance, in the SIRS model with secondary infections~\cite{kaklamanos2024geometric}, which can be recast in the form~\eqref{eq:ODE_i_hom} with $N=6$ compartments and $\mathcal{I}=\{S,\,I,\,T,\,P,\,Y,\,R\}$. Specifically, besides the compartments $i=S,\,I,\,R$, three additional compartments are introduced:
\begin{enumerate*}[label=(\roman*)]
\item $i=T$, comprising the individuals that are temporarily immune after recovering from a primary infection;
\item $i=P$, comprising partly susceptible individuals that lost their transient immunity after recovering from a primary infection;
\item and $i=Y$, comprising the individuals that undergo a secondary infection.
\end{enumerate*}
On the basis of the diagram of Figure~\ref{fig:SIRSsquared}, the ODE system~\eqref{eq:ODE_i_hom} particularises to
\begin{equation}
    \resizebox{.93\textwidth}{!}{$\displaystyle
    \begin{cases}
        \dfrac{\di N_S}{\di t}=r p(S|P)N_P-(1-r)\bigl(q(I|S,I)N_I+q(I|S,Y)N_Y\bigr)N_S, \\[3mm]
	\dfrac{\di N_I}{\di t}=-rp(T|I)N_I+(1-r)\bigl(q(I|S,I)N_I+q(I|S,Y)N_Y\bigr)N_S, \\[3mm]
	\dfrac{\di N_T}{\di t}=r\bigl(p(T|I)N_I-p(P|T)N_T\bigr), \\[3mm]
        \dfrac{\di N_P}{\di t}=r\bigl(p(P|T)N_T-p(S|P)N_P+p(P|R)N_R\bigr)-(1-r)\bigl(q(Y|P,I)N_I+q(Y|P,Y)N_Y\bigr)N_P, \\[3mm]
        \dfrac{\di N_Y}{\di t}=-rp(R|Y)N_Y+(1-r)\bigl(q(Y|P,I)N_I+q(Y|P,Y)N_Y\bigr)N_P, \\[3mm]
        \dfrac{\di N_R}{\di t}=r\bigl(p(R|Y)N_Y-p(P|R)N_R\bigr).
    \end{cases}
    $}
    \label{eq:SIRSsquared}
\end{equation}

Since both $i=I$ and $i=Y$ are infectious compartment, in this case the matrices defined via~\eqref{eq:MandV} are
$$  \mathbf{A}=
    \begin{pmatrix}
        (1-r)q(I|S,I) & (1-r)q(I|S,Y)\\
	0 & 0
    \end{pmatrix}, \qquad
    \mathbf{B}=
    \begin{pmatrix}
        rp(T|I) & 0 \\
	0 & rp(R|Y)
    \end{pmatrix}, $$
whence
$$  \mathbf{A}\mathbf{B}^{-1}=
    \begin{pmatrix}
        \frac{(1-r)q(I|S,I)}{rp(T|I)} & \frac{(1-r)q(I|S,Y)}{rp(R|Y)} \\
        0 & 0
    \end{pmatrix}, $$
and consequently, through~\eqref{eq:Rzero}, we find
$$ \mathcal{R}_0=\frac{(1-r)q(I|S,I)}{rp(T|I)}.$$
Notice that the above expression of $\mathcal{R}_0$ does not depend directly on any transition probability related to secondary infections.

Moreover, in this case the additional compartment-specific structuring variables can be interpreted as:
\begin{enumerate*}[label=(\roman*)]
    \item the level of \textit{temporary immunity} in compartment $i=T$;
    \item the level of \textit{partial immunity} in compartment $i=P$;
    \item the level of \textit{viral load} following a secondary infection in compartment $i=Y$.
\end{enumerate*}
The evolution of the mean values of the compartment-specific structuring variables is in this case governed by the following ODE system, which is derived from~\eqref{eq:ODE_M_i_hom}:
\begin{equation}
    \begin{cases}
	\dfrac{\di M_S}{\di t}=r\dfrac{p(S|P)(\bar{P}(S,P)-M_S)N_P}{N_S}, \\[3mm]
	\dfrac{\di M_I}{\di t}=(1-r)\dfrac{q(I|S,I)\bigl(\bar{Q}(I,S,I)-M_I\bigr)N_I+q(I|S,Y)\bigl(\bar{Q}(I,S,Y)-M_I\bigr)N_Y}{N_I}N_S, \\[3mm]
	\dfrac{\di M_T}{\di t}=r\dfrac{p(T|I)\bigl(\bar{P}(T,I)-M_T\bigr)}{N_T}N_I, \\[3mm]
	\dfrac{\di M_P}{\di t}=r\dfrac{p(P|T)(\bar{P}(P,T)-M_P)N_T+p(P|R)(\bar{P}(P,R)-M_P)N_R}{N_P}, \\[3mm]
	\dfrac{\di M_Y}{\di t}=(1-r)\dfrac{q(Y|P,I)\bigl(\bar{Q}(Y,P,I)-M_Y\bigr)N_I+q(Y|P,Y)\bigl(\bar{Q}(Y,P,Y)-M_Y\bigr)N_Y}{N_Y}N_P, \\[3mm]
	\dfrac{\di M_R}{\di t}=r\dfrac{p(R|Y)\bigl(\bar{P}(R,Y)-M_R\bigr)}{N_R}N_Y.
    \end{cases}
    \label{eq:exSIRSsquared_Mi}
\end{equation}

Backward bifurcation for $\mathcal{R}_0=1$ has been observed in many other epidemic models but mainly including demography~\cite{brauer2004backward,gumel2012causes,steindorf2022modeling,wangari2016backward}. The modelling framework presented here, which instead does not account for birth and death phenomena owing to global mass conservation (cf.~\eqref{ass:f}), can describe backward bifurcation as a consequence of simple compartment switching.

\begin{figure}[!t]
\centering
\subfigure[]{\includegraphics[width=.45\textwidth]{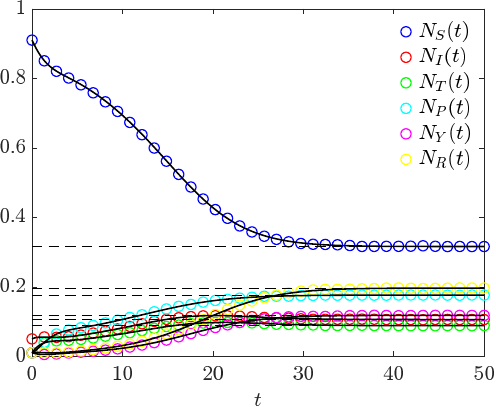}} \qquad
\subfigure[]{\includegraphics[width=.45\textwidth]{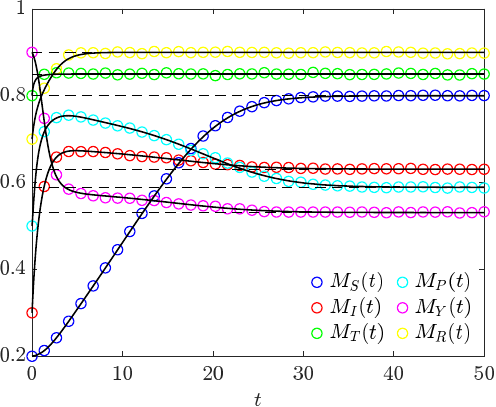}}
\caption{SIRS model with secondary infections. Solid lines: numerical solutions of~\eqref{eq:SIRSsquared} (panel (a)) and~\eqref{eq:exSIRSsquared_Mi} (panel (b)); circles: results of Monte Carlo simulations of the individual-based model governed by~\eqref{eq:microdynlaws} adapted to this case; dashed lines: analytical equilibria. Relevant parameters: $p(S|P)=p(P|R)=0.5p(P|T)$, $rp(P|T)=0.15$, $(1-r)q(I|S,I)=0.1225$, $q(I|S,Y)=2q(I|S,I)$, $rp(T|I)=0.125$, $q(Y|P,I)=2q(I|S,I)$, $rp(R|Y)=0.125$, $\bar{P}(S,P)=0.8$, $\bar{Q}(I,S,I)=0.7$, $\bar{Q}(I,S,Y)=0.6$, $\bar{P}(T,I)=0.85$, $\bar{P}(P,T)=0.8$, $\bar{P}(P,R)=0.4$, $\bar{Q}(Y,P,I)=0.6$, $\bar{Q}(Y,P,Y)=0.5$, $\bar{P}(R,Y)=0.9$. Hence, $\mathcal{R}_0=0.98$. Initial condition: $(N_S(0),\,N_I(0),\,N_T(0),\,N_P(0),\,N_Y(0),\,N_R(0),\,M_S(0),\,M_I(0),\,M_T(0),\,M_P(0),\,M_Y(0),\,M_R(0))=(0.91,\,0.05,\,0.01,\,0.01,\,0.01,\,0.01,\,0.2,\,0.3,\,0.8,\,0.5,\,0.9,\,0.7)$. Monte Carlo simulations are carried out using $10^6$ agents and defining the model parameters and functions as detailed in Appendix~\ref{sec:appibm2}}
\label{fig:SIRS2_simu}
\end{figure}

The plots in Figure~\ref{fig:SIRS2_simu} show a comparison between numerical solutions of the ODE systems~\eqref{eq:SIRSsquared} and \eqref{eq:exSIRSsquared_Mi} and the results of Monte Carlo simulations of the corresponding individual-based model, which indicate that convergence to the EE occurs even though $\mathcal{R}_0<1$.

\subsection{SIRWS model}\label{sec:SIRWS_ex}
We present now a model, also included in the framework~\eqref{eq:ODE_i_hom}, wherein interaction-driven compartmental switching does not lead individuals to enter an infectious compartment. Specifically, to consider the SIRWS model~\cite{dafilis2012influence,jardon2021geometric}, we choose $N=4$ and $\mathcal{I}=\{S,\,I,\,R,\,W\}$, where the compartment $i=W$ comprises individuals undergoing a ``waning immunity'' phase: while in this compartment, interacting with an infectious individual boosts the immunity, which induces a switch back to compartment $i=R$. Otherwise, an individual will eventually lose immunity completely and move to compartment $i=S$.

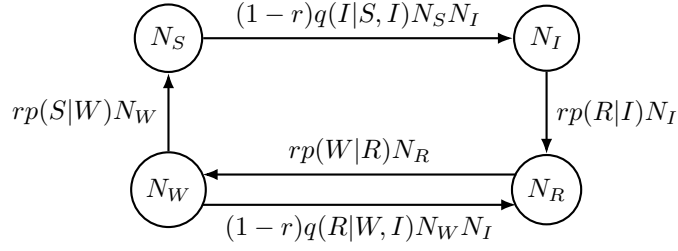
\begin{figure}[!t]
\centering
\begin{tikzpicture}
    \node[draw,circle,thick,minimum size=.75cm] (s) at (1,0) {$N_S$};
    \node[draw,circle,thick,minimum size=.75cm] (i) at (6,0) {$N_I$};
    \node[draw,circle,thick,minimum size=.75cm] (r) at (6,-2) {$N_R$};
    \node[draw,circle,thick,minimum size=.75cm] (w) at (1,-2) {$N_W$};
    \draw[-{Latex[length=2.mm, width=1.5mm]},thick] (s)--(i) node[above, midway]{$(1-r)q(I|S,I)N_SN_I$};
    \draw[-{Latex[length=2.mm, width=1.5mm]},thick] (i)--(r) node[right, midway]{$rp(R|I)N_I$};
    \draw[-{Latex[length=2.mm, width=1.5mm]},thick] (w)--(s) node[left,midway]{$rp(S|W)N_W$};
    \draw[{Latex[length=2.mm, width=1.5mm]}-,thick] ($(w)+(0.45,0.2)$)--($(r)+(-0.4,0.2)$) node[above,midway]{$rp(W|R)N_R$};
    \draw[-{Latex[length=2.mm, width=1.5mm]},thick] ($(w)+(0.45,-0.2)$)--($(r)+(-0.4,-0.2)$) node[below,midway]{$(1-r)q(R|W,I)N_W N_I$};
\end{tikzpicture}
\caption{Flow diagram of the SIRWS model}
\label{fig:SIRWS}
\end{figure}

In~\cite{dafilis2012influence,jardon2021geometric} the authors include demography, which we neglect here so as to recast their model in our framework. Following the diagram depicted in Figure~\ref{fig:SIRWS}, we derive from~\eqref{eq:ODE_i_hom} the following ODE system:
\begin{equation}
    \begin{cases}
        \dfrac{\di N_S}{\di t}=rp(S|W)N_W-(1-r)q(I|S,I)N_SN_I, \\[3mm]
	\dfrac{\di N_I}{\di t}=-rp(R|I)N_I+(1-r)q(I|S,I)N_SN_I, \\[3mm]
	\dfrac{\di N_R}{\di t}=r\bigl(p(R|I)N_I-p(W|R)N_R\bigr)+(1-r)q(R|W,I)N_WN_I, \\[3mm]
	\dfrac{\di N_W}{\di t}=r\bigl(p(W|R)N_R-p(S|W)N_W\bigr)-(1-r)q(R|W,I)N_WN_I.
    \end{cases}
    \label{eq:SIRWS}
\end{equation}
The matrices $\mathbf{A}$ and $\mathbf{B}$ of this system are again scalar quantities, as it is the case for the ODE system~\eqref{eq:exSIRS}, because only one infectious compartment is present, that is,
$$\mathbf{A} \equiv A=(1-r)q(I|S,I), \qquad \mathbf{B} \equiv B =r\bigl(1-p(I|I)\bigr)=rp(R|I). $$
The above expression for $B$ follows from~\eqref{eq:ker_assump5c} together with the fact that, in this model, $p(S|I)=p(W|I)=0$. Through~\eqref{eq:Rzero} we find 
$$ \mathcal{R}_0=\frac{A}{B}=\frac{(1-r)q(I|S,I)}{rp(R|I)}, $$
i.e. the same basic reproduction number as that of the SIRS model.

Moreover, in this case the additional compartment-specific structuring variable can be interpreted as the level of \textit{immunity} in compartment $i=W$. The evolution of the mean values of the compartment-specific structuring variables is now governed by the following ODE system, which is again derived from~\eqref{eq:ODE_M_i_hom}:
\begin{equation}
    \begin{cases}
	\dfrac{\di M_S}{\di t}=r\dfrac{p(S|W)(\bar{P}(S,W)-M_S)}{N_S}N_W, \\[3mm]
	\dfrac{\di M_I}{\di t}=(1-r)q(I|S,I)(\bar{Q}(I,S,I)-M_I)N_S, \\[3mm]
	\dfrac{\di M_R}{\di t}=\dfrac{rp(R|I)(\bar{P}(R,I)-M_R)+(1-r)q(R|W,I)(\bar{Q}(R,W,I)-M_R)N_W}{N_R}N_I, \\[3mm]
	\dfrac{\di M_W}{\di t}=r\dfrac{p(W|R)(\bar{P}(W,R)-M_W)}{N_W}N_R.
    \end{cases}
    \label{eq:exSIRWS_Mi}
\end{equation}
Assuming that the fractions of individuals in each compartment tend asymptotically in time to the EE, say $(N_S^\ast,\,N_I^\ast,\,N_R^\ast,\,N_W^\ast)$, we see from~\eqref{eq:exSIRWS_Mi} that the corresponding mean compartment-specific structuring variables tend to $M_S^\ast=\bar{P}(S,R)$, $M_I^\ast=\bar{Q}(I,S,I)$, $M_W^\ast=\bar{P}(W,R)$, and
$$ M_R^\ast=\frac{rp(R|I)\bar{P}(R,I)+(1-r)q(R|W,I)\bar{Q}(R,W,I)N_W^\ast}{rp(R|I)+(1-r)q(R|W,I)N_W^\ast}. $$
Since the SIRWS model admits periodic limit cycles, see \cite{dafilis2012influence,jardon2021geometric}, we might observe periodic fluctuations in $M_R$, which depends on $N_W$, while the mean values of all the other compartment-specific structuring variables, which are independent of the fractions of individuals in the various compartments, converge to an equilibrium. To better investigate this possibility, we look at the equilibrium value of $M_R$ as a function of $N_W$
$$ M_R(N_W)=\frac{rp(R|I)\bar{P}(R,I)+(1-r)q(R|W,I)\bar{Q}(R,W,I)N_W}{rp(R|I)+(1-r)q(R|W,I)N_W} $$
and we compute
$$ \frac{\di M_R}{\di N_W}=r(1-r)p(R|I)q(R|W,I)\frac{\bar{Q}(R,W,I)-\bar{P}(R,I)}{{\left(rp(R|I)+(1-r)q(R|W,I)N_W\right)}^2}, $$
whence we discover that $M_R$ is strictly increasing or decreasing with respect to $N_W$ depending on the sign of $\bar{Q}(R,W,I)-\bar{P}(R,I)$. In particular, $\bar{Q}(R,W,I)>\bar{P}(R,I)$ results in peaks of $M_R$ while $\bar{Q}(R,W,I)<\bar{P}(R,I)$ in dips of $M_R$. If instead $\bar{Q}(R,W,I)=\bar{P}(R,I)$ then $M_R$ does not oscillate. Moreover, large values of $\bar{Q}(R,W,I)-\bar{P}(R,I)$ result in large excursions of $M_R$ away from the equilibrium value $M_R^\ast$. Nevertheless, we observe that the term $N_I$ in the equation for $M_R$ in~\eqref{eq:exSIRWS_Mi} makes the evolution of $M_R$ quite slow if the epidemic is dormant, i.e. $N_I\approx 0$. Therefore, excursions away from $M_R^\ast$ can be clearly observed only when $N_I$ grows sufficiently far from $0$. 

\begin{figure}[!t]
\centering
\subfigure[]{\includegraphics[width=.45\textwidth]{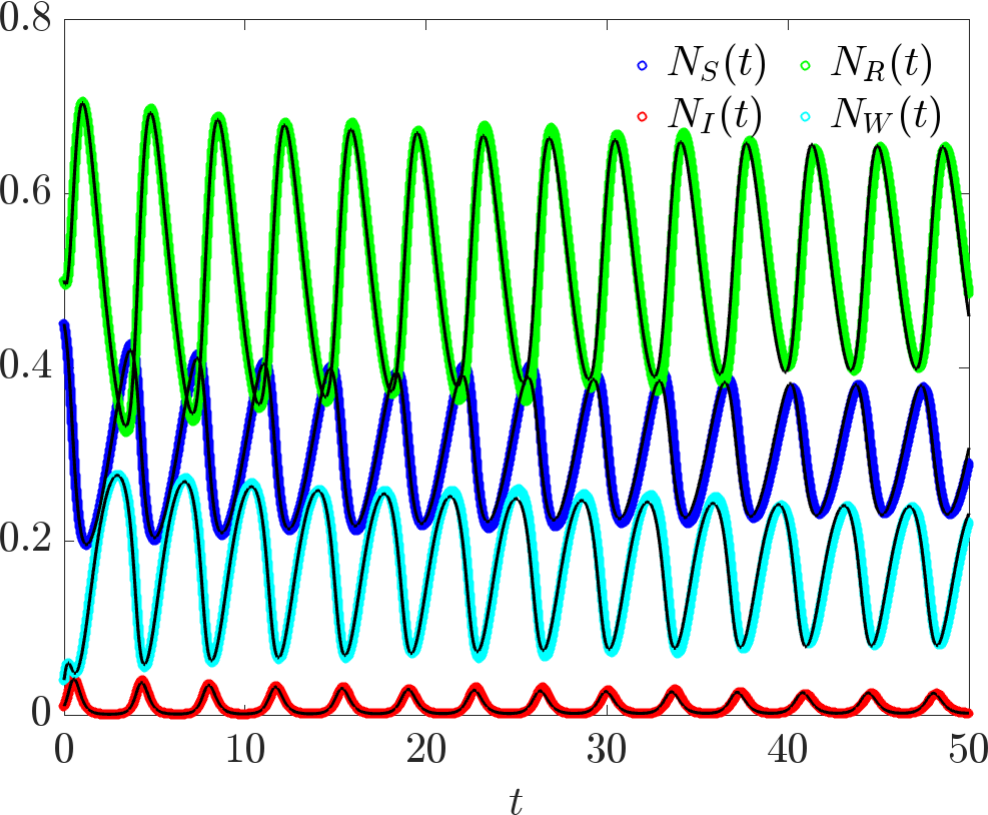}} \qquad
\subfigure[]{\includegraphics[width=.45\textwidth]{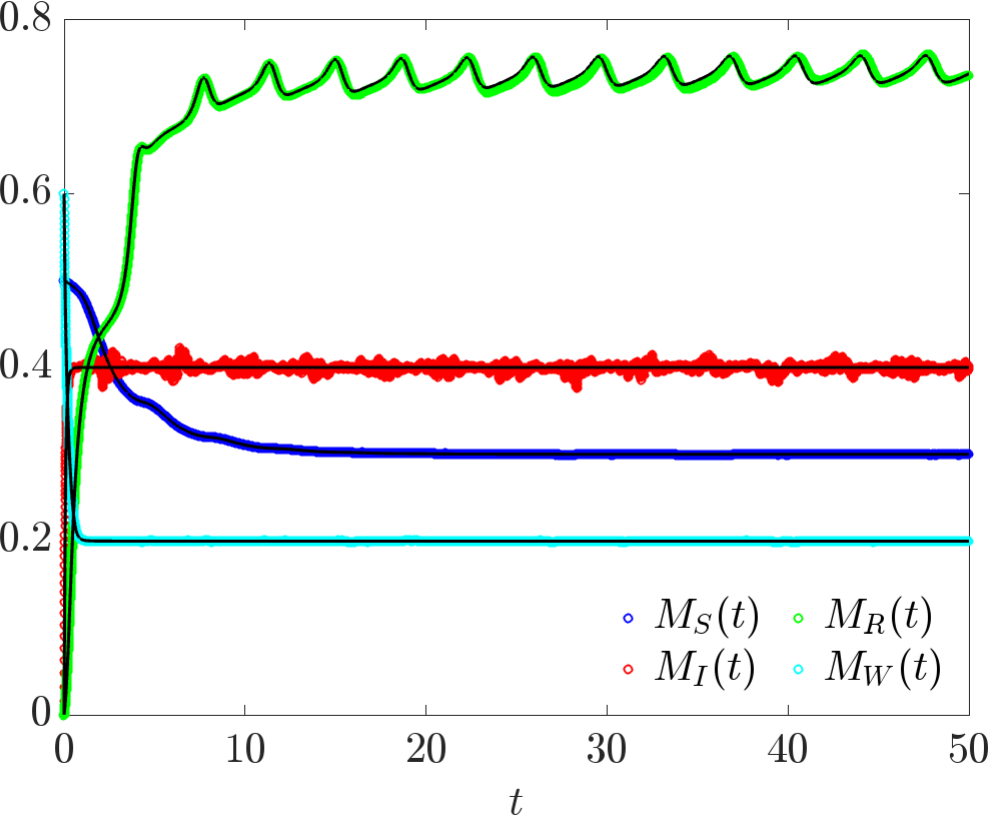}}
\caption{SIRWS model (without demography). Solid lines: numerical solutions of~\eqref{eq:SIRWS} (panel (a)) and~\eqref{eq:exSIRWS_Mi} (panel (b)); circles: results of Monte Carlo simulations of the individual-based model governed by~\eqref{eq:microdynlaws} adapted to this case. Relevant parameters: $rp(S|W)=rp(W|R)=1/3100$, $(1-r)q(I|S,I)=3/31$, $rp(R|I)=17/310$, $q(R|W,I)=10q(I|S,I)$, $(1-r)q(R|W,I)=30/31$, $\bar{P}(S,W)=0.3$, $\bar{Q}(I,S,I)=0.4$, $\bar{Q}(R,W,I)=0.9$, $\bar{P}(R,I)=0.4$, $\bar{P}(W,R)=0.2$. Hence, $\mathcal{R}_0=30/17$. Initial condition: $(N_S(0),\,N_I(0),\,N_R(0),\,N_W(0),\,M_S(0),\,M_I(0),\,M_R(0),\,M_W(0))=(0.97,\,0.01,\,0.01,\,0.01,\,0.5,\,0,\,0.6,\,0)$. Monte Carlo simulations are carried out using $10^6$ agents and defining the model parameters and functions as detailed in Appendix~\ref{sec:appibm3}}
\label{fig:SIRWS_simu}
\end{figure}

The plots in Figure~\ref{fig:SIRWS_simu} show a comparison between numerical solutions of the ODE systems~\eqref{eq:SIRWS} and \eqref{eq:exSIRWS_Mi} and the results of Monte Carlo simulations of the corresponding individual-based model. Notice how $M_R$ reacts to the dips of $N_W$, which coincide with the spikes of $N_I$ representing consequent epidemic waves. 

\section{Conclusions}\label{sec:concl}
We developed a general modelling framework for heterogeneously structured compartmental epidemiological systems. In this framework, each compartment, labelled by an index $i \in \mathcal{I}$, is structured by a specific continuous variable, $v_i \in \mathcal{V}_i$, which represents the level of expression of  a compartment-specific trait. The framework takes into account the effects of structuring-variable switching (i.e. the fact that the levels of expression of the compartment-specific traits can evolve in time) and compartment switching (i.e. the fact that individuals can transition between compartments), where the latter can be both spontaneous or driven by interactions amongst individuals belonging to the same or different compartments.

We first formulated a stochastic individual-based model that tracks the dynamics of single individuals, from which we formally derived the corresponding mesoscopic model, which consists of the IDE system~\eqref{eq:IDE_i} for the population density functions, $n_i(t,v_i)$, of the various compartments at time $t$. We then considered an appropriately rescaled version of this system, which is given by the IDE system~\eqref{eq:IDE_i_resc1}, and we carried out formal asymptotic analyses to derive the corresponding macroscopic model.

The macroscopic model so obtained comprises the ODE system~\eqref{eq:ODE_Ni_01},\eqref{eq:ODE_NiMi_01} for the fractions, $N_i(t)$, of individuals in the different compartments and the mean values, $M_i(t)$, of the compart\-ment-specific structuring variables. In this ODE system, the dynamics of such macroscopic quantities are described in terms of the microscopic information contained in the model parameters and functions. From the ODE system~\eqref{eq:ODE_Ni_01},\eqref{eq:ODE_NiMi_01}, under the homogeneity assumption that the model functions depend on the compartment indices but are independent of the structuring variables, we obtained the reduced ODE systems~\eqref{eq:ODE_i_hom} and~\eqref{eq:ODE_M_i_hom}, and then, employing the Next Generation Matrix approach, we obtained a general formula for the basic reproduction number, $\mathcal{R}_0$, in terms of key parameters and functions of the underlying microscopic model. This illustrates how the modelling framework developed here makes it possible to draw connections between fundamental individual-level processes and population-scale dynamics. 

Moreover, we applied the modelling framework to case studies from classical compartmental epidemiological systems (i.e. the SIRS model, the SIRS model with secondary infections, and the SIRWS model) and, for each of them, we showed that there is excellent agreement between the results of Monte Carlo simulations of the individual-based model and both numerical solutions and analytical results of the macroscopic model. This validates the formal limiting procedures employed to obtain the mesoscopic and macroscopic models from the underlying individual-based model. 

Possible extensions of the present work include incorporating demography into the modelling framework, in order to take into account proliferation and death of individuals in one or more compartments. This would require generalising both the individual-based modelling approach as well as the formal approach employed in this paper to derive the corresponding mesoscopic and macroscopic models, since both approaches rely on the fact that the total number of individuals in the system is conserved, which would not be the case if demography was incorporated. In the same vein, building for instance on the methods employed in~\cite{cristiani2025kinetic,estrada2023macroscopic,lorenzi2024phenotype} in different application domains, an additional development of this work would be to extend the individual-based modelling approach and the derivation methods employed here to the case where a spatial structure is also included, in order to then investigate how the interplay between spatial movement and both structuring-variable switching and compartment switching may impact on the evolutionary dynamics of epidemiological systems.

Moreover, while the modelling approach presented here is framed in an abstract context to highlight general properties, it would be interesting to apply it to the study of the dynamics of specific infectious diseases. In particular, an additional avenue for future research would be to explore the possibility to estimate the forms of the model functions based on data; for this, techniques similar to those employed in~\cite{albi2022kinetic,albi2021control,dimarco2021kinetic} may prove useful. \red{This would enable us to describe specific dynamics of real-world diseases; for instance, we could incorporate the possibility of quarantining an infectious individual with a high viral load by setting their probability of infecting other individuals to zero. This could be of particular relevance for the recent COVID-19 pandemic, and could also be of interest in general for transmissible diseases for which quarantine is a feasible and functioning form of control.}

Finally, while our attention has been focused on the microscopic and macroscopic models, it would be relevant to consider also the mesoscopic model defined by the IDE system~\eqref{eq:IDE_i}. In this regard, it would be interesting to investigate possible limiting regimes and conditions on the model terms under which the population density functions become unimodal or multimodal (i.e. the various compartments become monomorphic or polymorphic). For this, techniques similar to those employed in~\cite{alfaro2018evolutionary,desvillettes2008selection,Jabin2011,lorenzi2020asymptotic,Perthame2006,Pouchol2017b} may prove useful. This would allow for further investigation into the mechanisms and processes underpinning the emergence of inter-individual and intra-compartmental heterogeneity in epidemiological systems.  

\appendix

\section{Proofs of~\texorpdfstring{\eqref{eq:asyres1},~\eqref{eq:asyres3}}{}}
\label{appendix:A}
Recall that the $\Theta_i$'s, $i\in \mathcal{I}$, are independent Bernoulli random variables with parameters $\theta_i\Delta{t}$. We introduce the random variable 
$$\mathfrak{S}:=\sum_{i\in\mathcal{I}}\Theta_i, $$
and we want to show that, for $\Delta{t}$ small, it holds that ${\rm P}(\mathfrak{S}\geq 2)=o(\Delta t)$. Clearly,
$$ {\rm P}(\mathfrak{S}\geq 2)=1-\bigl({\rm P}(\mathfrak{S}=0)+{\rm P}(\mathfrak{S}=1)\bigr) $$
and, furthermore,
$$ {\rm P}(\mathfrak{S}=0)={\rm P}(\Theta_i=0\ \forall\,i\in\mathcal{I})=\prod_{i\in\mathcal{I}}(1-\theta_i\Delta{t}), \qquad
    {\rm P}(\mathfrak{S}=1)=\sum_{i\in\mathcal{I}}\theta_i\Delta{t}\prod_{\substack{j\in\mathcal{I} \\ j\neq i}}(1-\theta_j\Delta{t}). $$
Now, let $\displaystyle{p:=\prod_{i\in \mathcal{I}} (1-\theta_i \Delta t)}$; then
$$ \log{p}=\sum_{i\in\mathcal{I}}\log{(1-\theta_i\Delta{t})}=
    -\sum_{i\in\mathcal{I}}\theta_i\Delta{t}+o(\Delta{t}) \quad \text{as} \quad \Delta{t}\to 0^+, $$
whence 
$$ p=e^{\log{p}}=e^{-\sum_{i\in\mathcal{I}}\theta_i\Delta{t}+o(\Delta{t})}
    =1-\sum_{i\in\mathcal{I}}\theta_i\Delta{t}+o(\Delta{t}) \quad \text{as} \quad \Delta{t}\to 0^+. $$
Using this, we can also write
\begin{align*}
    {\rm P}(\mathfrak{S}=1) &= \sum_{i\in\mathcal{I}}\theta_i\Delta{t}\left(1-\sum_{j\neq i}\theta_j\Delta{t}+o(\Delta{t})\right) \\
    &= \sum_{i\in\mathcal{I}}\theta_i\Delta{t}-\sum_{i\in\mathcal{I}}\sum_{\substack{j\in\mathcal{I} \\ j\neq i}}\theta_i\theta_j\Delta{t}^2
        +o(\Delta{t}^2) \\
    &= \sum_{i\in\mathcal{I}}\theta_i\Delta{t}+o(\Delta{t}^2)
\end{align*}
and finally
$$ {\rm P}(\mathfrak{S}\geq 2)=1-\left(1-\sum_{i\in\mathcal{I}}\theta_i\Delta{t}+o(\Delta{t})
+\sum_{i\in\mathcal{I}}\theta_i\Delta{t}+o(\Delta{t}^2)\right)=o(\Delta{t}) \quad \text{as} \quad \Delta{t}\to 0^+, $$
which proves the claim.

\section{Definitions of the parameters and functions used in simulations of the individual-based models}
\label{sec:appibm}

\subsection{SIRS model}
\label{sec:appibm1}
The model parameters and functions are defined as follows:

$$
\Delta t =10^{-2}, \quad r=0.5, \quad \theta_i = 0.5\;\; \forall \, i \in \mathcal{I},
$$

$$
K_i(v_i'|v_i) \coloneqq \delta(v_i'-v_i)\;\; \forall \, i \in \mathcal{I},
$$

$$
p(j|i,v_i)\equiv p(j|i) \coloneqq \begin{cases}
  \delta_{jS} &\quad\text{for}\quad i=S, \\
 0.2\,\delta_{jR}+ 0.8\,\delta_{jI} &\quad\text{for}\quad i=I, \\
 0.1\,\delta_{jS}+0.9\,\delta_{jR} &\quad\text{for}\quad i=R,
\end{cases}
$$

$$
q(j|i,v_i,k,v_k^\ast)\equiv q(j|i,k) \coloneqq  \begin{cases}
 0.4\,\delta_{jI} + 0.6\,\delta_{jS}&\quad\text{for}\quad i=S,\; k=I, \\
 \delta_{ji}&\quad\text{otherwise.}
\end{cases}
$$

Moreover, the terms $P(v_j''|j,i,v_i)\equiv P(v_j''|j,i)$ and $Q(v_j''|j,i,v_i,k,v_k^\ast)\equiv Q(v_j''|j,i,k)$ which are relevant (i.e. either different from zero or with $j\neq i$) for the present case study are defined as

$$
P(v_j''|j,i) \coloneqq \begin{cases}
    \mathbbm{1}_{[0,\,1]}(v_j''),&\quad\text{for}\quad j=R,\;i=I, \\
    \frac{5}{3}\mathbbm{1}_{[0,\,0.6]}(v_j'')&\quad\text{for}\quad j=S,\;i=R, 
\end{cases} 
$$

$$ Q(v_j''|j,i,k) \coloneqq \frac{5}{4}\mathbbm{1}_{[0,\,0.8]}(v_j'') \quad \text{for} \quad j=k=I,\;i=S. $$

At the initial time $t=0$, the individuals in the various compartments are distributed according to the following distributions:
$$
n_i(0,v_i)\coloneqq \frac{N_i(0)}{2M_i(0)}\mathbbm{1}_{[0,\,2M_i(0)]}(v_i), \qquad i\in\mathcal{I},
$$
where the values of $N_i(0)$ and $M_i(0)$ are reported in the caption of Figure~\ref{fig:SIRS_simu}.

\subsection{SIRS model with secondary infection}
\label{sec:appibm2}

The model parameters and functions are defined as follows:

$$
\Delta t =10^{-2}, \quad r=0.5, \quad \theta_i = 0.5\;\; \forall \, i \in \mathcal{I},
$$

$$
K_i(v_i'|v_i) \coloneqq \delta(v_i'-v_i)\;\; \forall \, i \in \mathcal{I},
$$

$$
p(j|i,v_i)\equiv p(j|i) \coloneqq \begin{cases}
 \delta_{jS}   &\quad\text{for}\quad i=S\\
 0.25\,\delta_{jT} + 0.75\,\delta_{jI} &\quad\text{for}\quad i=I, \\
 0.30\,\delta_{jP} + 0.70\,\delta_{jT} &\quad\text{for}\quad i=T, \\
 0.15\,\delta_{jS} + 0.85\,\delta_{jP} &\quad\text{for} \quad i=P, \\
 0.25\,\delta_{jR} + 0.75\,\delta_{jY} &\quad\text{for}\quad i=Y, \\
 0.15\,\delta_{jP} + 0.85\,\delta_{jR} &\quad\text{for}\quad i=R,
\end{cases}
$$

$$
q(j|i,v_i,k,v_k^\ast)\equiv q(j|i,k) \coloneqq  \begin{cases}
 0.245\delta_{jI}+0.755\delta_{jS} & \quad\text{for}\quad i=S,\; k=I, \\
 0.490\delta_{jI}+0.510\delta_{jS} &\quad\text{for}\quad i=S,\;k=Y, \\
 0.490\delta_{jY}+0.510\delta_{jP} &\quad\text{for}\quad i=P,\;k=I, \\
 0.980\delta_{jY}+0.020\delta_{jP} &\quad\text{for}\quad i=P,\;k=Y, \\ 
 \delta_{ji}&\quad\text{otherwise.}
\end{cases}
$$

Moreover, the terms $P(v_j''|j,i,v_i)\equiv P(v_j''|j,i)$ and $Q(v_j''|j,i,v_i,k,v_k^\ast)\equiv Q(v_j''|j,i,k)$ which are relevant (i.e. either different from zero or with $j\neq i$) for the present case study are defined as

$$
P(v_j''|j,i) \coloneqq \begin{cases}
  \frac{10}{17} \mathbbm{1}_{[0,\,1.7]}(v_j'') &\quad\text{for}\quad j=T,\;i=I,\\
  \frac{5}{8}\mathbbm{1}_{[0,\,1.6]}(v_j'') &\quad\text{for}\quad j=P,\;i=T, \\
  \frac{5}{8}\mathbbm{1}_{[0,\,1.6]}(v_j'') &\quad\text{for}\quad j=S,\;i=P,\\
  \frac{5}{9}\mathbbm{1}_{[0,\,1.8]}(v_j'') &\quad\text{for}\quad j=R,\;i=Y, \\
 \frac{5}{4}\mathbbm{1}_{[0,\,0.8]} 
 (v_j'') &\quad\text{for}\quad j=P,\;i=R,   
\end{cases}
$$

$$
Q(v_j''|j,i,k) \coloneqq \begin{cases}
 \frac{5}{7} \mathbbm{1}_{[0,\,1.4]}(v_j'') &\quad\text{for}\quad j=k=I,\;i=S, \\
 \frac{5}{6} \mathbbm{1}_{[0,\,1.2]}(v_j'') &\quad\text{for}\quad j=I,\;i=S,\;k=Y, \\
 \frac{5}{6} \mathbbm{1}_{[0,\,1.2]}(v_j'') &\quad\text{for}\quad j=Y,\;i=P,\;k=I, \\
 \mathbbm{1}_{[0,\,1]}(v_j'') &\quad\text{for}\quad j=k=Y,\;i=P.
\end{cases}
$$

At the initial time $t=0$, the individuals in the various compartments are distributed according to the following distributions:
$$
n_i(0,v_i)\coloneqq \frac{N_i(0)}{2M_i(0)}\mathbbm{1}_{[0,\,2M_i(0)]}(v_i), \qquad i\in\mathcal{I},
$$
where the values of $N_i(0)$ and $M_i(0)$ are reported in the caption of Figure~\ref{fig:SIRS2_simu}.

\subsection{SIRWS model}
\label{sec:appibm3}
The model parameters and functions are defined as follows:

$$
\Delta t =0.0032, \quad r=0.05, \quad \theta_i = 0.5\;\; \forall \, i \in \mathcal{I},
$$

$$
K_i(v_i'|v_i) \coloneqq \delta(v_i'-v_i)\;\; \forall \, i \in \mathcal{I}, \;\; \forall \, i \in \mathcal{I},
$$

$$
p(j|i,v_i)\equiv p(j|i) \coloneqq \begin{cases}
  \delta_{ji} &\quad\text{for}\quad i=S,\\
 0.5806\,\delta_{jR}+0.4194\,\delta_{jI} &\quad\text{for}\quad i=I, \\
 0.0323\,\delta_{jW}+0.9677\,\delta_{jR} &\quad\text{for}\quad i=R, \\
 0.0323\,\delta_{jS}+0.9677\,\delta_{jW} &\quad\text{for}\quad i=W, 
\end{cases}
$$

$$
q(j|i,v_i,k,v_k^\ast)\equiv q(j|i,k) \coloneqq  \begin{cases}
 0.1019\,\delta_{jI} + 0.8981\,\delta_{jS} &\quad\text{for}\quad i=S,\; k=I, \\
 0.5093\,\delta_{jR} + 0.4907\,\delta_{jW} &\quad\text{for}\quad i=W,\; k=I, \\
 \delta_{ji} &\quad\text{otherwise.}
\end{cases}
$$

Moreover, the terms $P(v_j''|j,i,v_i)\equiv P(v_j''|j,i)$ and $Q(v_j''|j,i,v_i,k,v_k^\ast)\equiv Q(v_j''|j,i,k)$ which are relevant (i.e. either different from zero or with $j\neq i$) for the present case study are defined as

$$
P(v_j''|j,i) \coloneqq \begin{cases}
 \frac{5}{4} \mathbbm{1}_{[0,\,0.8]}(v_j'')  &\quad\text{for}\quad j=R,\;i=I, \\
 \frac{5}{2} \mathbbm{1}_{[0,\,0.4]}(v_j'') &\quad\text{for}\quad j=W,\;i=R, \\
 \frac{5}{3} \mathbbm{1}_{[0,\,0.6]}(v_j'') &\quad\text{for}\quad j=S,\;i=W, 
\end{cases}
$$

$$
Q(v_j''|j,i,k) \coloneqq  \begin{cases}
 \frac{5}{4} \mathbbm{1}_{[0,\,0.8]}(v_j'') &\quad\text{for}\quad j=k=I,\;i=S,\\
 \frac{5}{9} \mathbbm{1}_{[0,\,1.8]}(v_j'') &\quad\text{for}\quad j=R,\;i=W,\; k=I.
\end{cases}
$$

At the initial time $t=0$, the individuals in the various compartments are distributed according to the following distributions:
$$
n_i(0,v_i)\coloneqq \frac{N_i(0)}{2M_i(0)}\mathbbm{1}_{[0,\,2M_i(0)]}(v_i), \qquad i\in\mathcal{I},
$$
where the values of $N_i(0)$ and $M_i(0)$ are reported in the caption of Figure~\ref{fig:SIRWS_simu}.

\bibliographystyle{plain}
\bibliography{biblio}
\end{document}